\pdfoutput=1

\documentclass[11pt,a4paper]{article}
\usepackage{bm}
\usepackage{jheppub}
\usepackage{bbm}
\usepackage{mathrsfs}

\allowdisplaybreaks
\let\de=\partial
\let\eps=\epsilon
\let\a=\alpha
\let\b=\beta
\let\g=\gamma
\let\d=\delta
\let\k=\kappa
\let\l=\lambda
\let\m=\mu
\let\n=\nu
\let\s=\eps
\let\t=\theta
\let\om=\omega
\let\Om=\Omega
\let\x=\xi
\let\w=\wedge

\newcommand\im{\text{i}}
\newcommand\gr[1]{\mathrm{#1}}
\newcommand\La{\mathscr{L}}
\newcommand\dd{\text{d}}
\newcommand\Ip{\amalg}
\DeclareMathOperator{\Co}{\widehat{ch}}
\DeclareMathOperator{\Si}{\widehat{sh}}
\DeclareMathOperator{\ho}{\star}
\newcommand\gal{\mathfrak{g}}
\newcommand\hal{\mathfrak{h}}
\newcommand\kal{\mathfrak{k}}
\newcommand{\vek}[1]{\bm{#1}}
\newcommand{\R}{\mathbb{R}}

\definecolor{myred}{rgb}{1,0,0}
\definecolor{myblue}{rgb}{0,0,1}
\newcommand\red[1]{{\color{myred}#1}}
\newcommand\blue[1]{{\color{myblue}#1}}

\title{Exceptional nonrelativistic effective field theories\\with enhanced symmetries}

\author{Tom\'a\v{s} Brauner}
\affiliation{Department of Mathematics and Physics, University of Stavanger,\\
N-4036 Stavanger, Norway}
\emailAdd{tomas.brauner@uis.no}

\abstract{We initiate the classification of nonrelativistic effective field theories (EFTs) for Nambu-Gold\-sto\-ne (NG) bosons, possessing a set of redundant, coordinate-dependent symmetries. Similarly to the relativistic case, such EFTs are natural candidates for ``exceptional'' theories, whose scattering amplitudes feature an enhanced soft limit, that is, scale with a higher power of momentum at long wavelengths than expected based on the mere presence of Adler's zero. The starting point of our framework is the assumption of invariance under spacetime translations and spatial rotations. The setup is nevertheless general enough to accommodate a variety of nontrivial kinematical algebras, including the Poincar\'e, Galilei (or Bargmann) and Carroll algebras. Our main result is an explicit construction of the nonrelativistic versions of two infinite classes of exceptional theories: the multi-Galileon and the multi-flavor Dirac-Born-Infeld (DBI) theories. In both cases, we uncover novel Wess-Zumino terms, not present in their relativistic counterparts, realizing nontrivially the shift symmetries acting on the NG fields. We demonstrate how the symmetries of the Galileon and DBI theories can be made compatible with a nonrelativistic, quadratic dispersion relation of (some of) the NG modes.}


\begin{document}
 
\maketitle


\section{Introduction}
\label{sec:intro}

Effective field theory (EFT) is a general framework that allows one to focus on the physics relevant at a given energy scale, and to dispense with irrelevant microscopic details. This approach is especially powerful in case of physical systems, possessing an ordered ground state, where the low-energy physics is dominated by collective modes: the Nambu-Goldstone (NG) bosons of the symmetry spontaneously broken by the order parameter. In case of spontaneously broken internal, coordinate-independent symmetries, the construction of EFT for NG bosons is by now well-understood in both relativistic~\cite{Coleman1969a,Callan1969a,Leutwyler1994b} and nonrelativistic~\cite{Leutwyler1994a,Watanabe2014a,Andersen2014a} systems. It has provided key insight into issues such as the geometry of spontaneous symmetry breaking, coupling of NG bosons to external fields or other dynamical degrees of freedom, and the counting of NG bosons (see refs.~\cite{Brauner2010a,Watanabe2020a,Beekman2019a,AlvarezGaume2020a} for a review of the latter subject).

The case of spontaneously broken spacetime, or more generally coordinate-dependent, symmetries is considerably more subtle. The most striking difference to coordinate-inde\-pen\-dent symmetries is that some broken symmetries now need not give rise to a NG boson~\cite{Ivanov1975a,Low2002a}. This issue is unrelated to possible canonical conjugation of different NG modes~\cite{Nambu2004a}, and arises from the fact that different symmetries may be locally indistinguishable~\cite{Watanabe2013a,Brauner2014a,Brauner2020a}. It is then natural to ask what physical consequences, if not the existence of NG modes in the spectrum, such ``redundant'' symmetries have.

A remarkable answer to this question was given in ref.~\cite{Cheung2015a}. Namely, another generic consequence of spontaneous symmetry breaking besides the very existence of NG bosons is that they interact weakly at low energies (except when they do not~\cite{Watanabe2014b,Rothstein2018a,Kampf2020a}). This means as a rule that in the long-wavelength, ``soft'' limit, scattering amplitudes of a NG boson are proportional to its momentum, a fact usually referred to as Adler's zero. It was shown in ref.~\cite{Cheung2015a} that the presence of redundant symmetries in the system leads to an enhancement of the soft limit, whereby the scattering amplitudes of NG bosons scale with some higher power of momentum. As of the time of writing this paper, no examples of theories with scattering amplitudes with enhanced soft limit \emph{not} caused by an underlying redundant symmetry are known. It is therefore reasonable to use symmetry as a starting point when searching for new interesting theories with nontrivial soft limits.

It turns out that the requirement that the soft limit of scattering amplitudes be enhanced restricts the pool of possible EFTs so strongly that their exhaustive classification seems feasible~\cite{Cheung2015a,Cheung2016a,Cheung2017a}. This was further supported by our previous work~\cite{Bogers2018a,Bogers2018b} where we carried out a direct classification of possible Lie algebra structures admitting redundant symmetries.\footnote{See also ref.~\cite{Roest2019a}, which somewhat systematized the approach put forward in refs.~\cite{Bogers2018a,Bogers2018b}, originally based on a brute-force enumeration of all constraints from Jacobi identities and their subsequent solution, and applied it to theories with higher-spin NG fields.} Among all relativistic EFTs with a \emph{single} NG boson, only two turn out to possess scattering amplitudes that display a soft limit enhanced beyond the expected Adler zero. Both of these enjoy a prominent position in particle physics as well as cosmology: the theory of fluctuations of a $D$-dimensional brane embedded into a $(D+1)$-dimensional pseudo-Euclidean spacetime, referred to as the Dirac-Born-Infeld (DBI) theory, and the Galileon theory~\cite{Nicolis2009a}.\footnote{While the DBI theory is completely determined by spacetime geometry, there are altogether $D+1$ different Galileon Lagrangians in $D$ spacetime dimensions. One of these is a tadpole that is undesirable in perturbative EFTs, and another is the usual kinetic term of the NG mode. This leaves us with $D-1$ possible interaction terms, further related by a set of dualities~\cite{Rham2014a,Rham2014b,Kampf2014a}.} An advantage of the Lie-algebraic approach is that it allows for a unified treatment of all EFTs regardless of the number of NG degrees of freedom (flavors). This led us in refs.~\cite{Bogers2018a,Bogers2018b} to the construction of new multi-Galileon and multi-flavor DBI theories where NG bosons whose scattering amplitudes possess enhanced soft limits are coupled to other, ``ordinary'' NG modes. This delimited the playground for studying EFTs with nontrivial scattering amplitudes and/or extended symmetries.

The goal of the present paper is to extend the analysis of refs.~\cite{Bogers2018a,Bogers2018b} to nonrelativistic theories, that is, theories lacking Lorentz invariance. The motivation for doing so is twofold. First, the world of quantum many-body systems --- whether relativistic systems at finite density or genuinely nonrelativistic, condensed-matter systems --- is rich in phenomenology, including nontrivial realization of spontaneously broken symmetries, that might otherwise be forbidden by Lorentz invariance. Second, non-Lorentzian kinematical algebras appear naturally in models inspired by nonrelativistic gravity~\cite{Horava2009a,Horava2009b} and nonrelativistic string theory, see e.g.~refs.~\cite{Andringa2012a,Harmark2017a,Kluson2018a}. We shall \emph{not} discuss the properties of scattering amplitudes of NG bosons in nonrelativistic EFTs here, the reason being that techniques for calculation of scattering amplitudes in such nonrelativistic systems, beyond the ordinary diagrammatic perturbation theory, have not been developed yet.\footnote{Scattering amplitudes of gapless particles in theories violating Lorentz invariance were addressed recently in ref.~\cite{Pajer2020a}. The discussion was however limited to theories where all particles have a strictly linear dispersion relation with the same phase velocity.} Until the precise correspondence between redundant symmetries and enhancement of soft limits of scattering amplitudes has been established, the results of this paper should therefore be considered as providing a list of \emph{candidate} theories where nontrivial scattering amplitudes might be observed.

The paper is organized as follows. In section~\ref{sec:algebra} we show that the results of refs.~\cite{Bogers2018a,Bogers2018b} can be applied verbatim to the nonrelativistic case. The material of this section can be viewed purely mathematically. Regardless of the above-introduced issues of spontaneous breaking of redundant symmetries and soft limits of scattering amplitudes, it simply lists possible consistent Lie algebras built upon the algebra of spacetime translations and spatial rotations by adding a further unspecified set of scalar and vector generators, the only additional assumption being that the scalar generators commute with \emph{spatial} translations. It thus extends some of the allowed nonrelativistic kinematic algebras~\cite{Bacry1968a,FigueroaOFarrill2017a,FigueroaOFarrill2018a}. As soon as one wants to turn a given Lie algebra into an actual low-energy EFT, one has to make a set of choices. These include the choice of physical NG degrees of freedom and the associated spontaneously broken symmetries, and the choice to add, if desired, certain redundant symmetries to the theory. The question for which of the allowed Lie algebras a perturbatively well-defined EFT actually exists, seems difficult to answer in full generality, although some partial steps were made in ref.~\cite{Bogers2018a}. We will therefore focus on constructing explicit examples of consistent EFTs. Two infinite classes of such examples constitute the material of sections~\ref{sec:galileon} and~\ref{sec:DBI}. In section~\ref{sec:summary} we summarize and conclude.

Some technical details that supplement the main text of the paper are relegated to three appendices. The class of symmetry Lie algebras presented in section~\ref{sec:algebra} was obtained using a general ansatz for the commutation relations, wherein only those contributions to the commutators that exist in an arbitrary number $d$ of spatial dimensions were included. In appendix~\ref{app:3d} we show that with a single additional, mild technical assumption, this ansatz is justified in the physically most interesting case of $d=3$. A discussion of the possibility of a presence of central charges in the Lie algebra of symmetry generators is provided in appendix~\ref{app:central}. Finally, in appendix~\ref{app:1flavor} we show that a complete classification of perturbatively well-defined EFTs based on the Lie algebras constructed in section~\ref{sec:algebra} is possible in the special case of a single physical NG degree of freedom.


\section{Extended symmetry algebra}
\label{sec:algebra}

In this section we will answer the purely mathematical question as to what symmetry structures can be obtained by augmenting the Lie algebra of spacetime translations and spatial rotations with additional, scalar or vector, generators. Only once the most general admissible Lie algebra of symmetry generators is at hand, will we comment on its possible physical interpretations.

We start by fixing the notation for the generators of spatial rotations, $J_{\m\n}$, and translations, $P_\m$. We assume these to span the Lie algebra of isometries of the Euclidean space,
\begin{align}
\notag
[J_{\m\n},J_{\k\l}]&=\im(g_{\m\l}J_{\n\k}+g_{\n\k}J_{\m\l}-g_{\m\k}J_{\n\l}-g_{\n\l}J_{\m\k}),\\
\label{euclidean}
[J_{\m\n},P_\l]&=\im(g_{\n\l}P_\m-g_{\m\l}P_\n),\\
\notag
[P_\m,P_\n]&=0.
\end{align}
Two remarks are in order here. First, in order to keep in touch with the notation introduced in refs.~\cite{Bogers2018a,Bogers2018b}, we break the usual convention and use Greek letters $\m,\n,\dotsc$ to denote \emph{spatial} indices. Whenever spacetime indices are needed later on, capital Latin letters, $M,N,\dotsc$, will be used. Second, $g_{\m\n}$ in eq.~\eqref{euclidean} stands for the positive-definite Euclidean metric. Throughout the paper, we will have implicitly in mind the standard Cartesian coordinates in which $g_{\m\n}=\delta_{\m\n}$. In order for eq.~\eqref{euclidean} to match the commutation relations between momentum and angular momentum, known from quantum mechanics, one should interpret $J_{\m\n}$ as minus the operator of angular momentum.

We would now like to add a set of additional scalar generators, $Q_i$, and additional vector generators, $K_{\m A}$. Rotation invariance restricts their commutators as follows,
\begin{align}
\notag
[J_{\m\n},K_{\l A}]&=\im(g_{\n\l}K_{\m A}-g_{\m\l}K_{\n A}),\\
\label{QKcomm}
[J_{\m\n},Q_i]&=0,\\
\notag
[Q_i,Q_j]&=\im f^k_{ij}Q_k,
\end{align}
where $f^k_{ij}$ are the structure constants of the Lie algebra of the scalars. The only additional assumption we will make is that the scalar generators $Q_i$ commute with spatial translations,
\begin{equation}
[P_\m,Q_i]=0.
\label{adler}
\end{equation}
The physical motivation behind this assumption is simply that it holds for most scalar symmetries in physics, and in case it does not --- as for instance for spatial dilatations --- the Adler zero property tends to be violated~\cite{Watanabe2014b,Rothstein2018a}. But since we are not attaching any physical interpretation to the Lie algebra under construction at this stage, we can just think of eq.~\eqref{adler} as a technical assumption that simplifies the analysis.

We can now copy-paste the result of ref.~\cite{Bogers2018a} that characterizes the most general Lie algebra satisfying the above constraints.\footnote{The problem addressed in ref.~\cite{Bogers2018a} was to find extensions of the Poincar\'e rather than the Euclidean algebra, but thanks to the fact that it was formulated solely in terms of the metric $g_{\m\n}$, the result can be transferred without change to the present nonrelativistic situation.} To that end, we introduce two (not necessarily disjoint) subsets of the scalar generators, $Q_A$ and $Q_{AB}$. The indices $A,B$ on these run over the same values as the capital Latin index on $K_{\m A}$, and moreover $Q_{AB}$ is required to be antisymmetric in its indices. It is, however, not necessary that all the $Q_A$s or $Q_{AB}$s are linearly independent or even nonzero. Furthermore, denoting the number of vectors $K_{\m A}$ as $n$, we introduce two $(n+1)\times(n+1)$ block matrices,
\begin{equation}
\arraycolsep=1ex
\renewcommand{\arraystretch}{1.5}
(T_i)^A_{\phantom AB}\equiv
\left(\begin{array}{c|c}
(t_i)^A_{\phantom AB} & 0\\
\hline
d_{Bi} & 0
\end{array}\right),\qquad
L_{AB}\equiv
\left(\begin{array}{c|c}
Q_{AB} & \im Q_A\\
\hline
-\im Q_B & 0
\end{array}\right).
\label{block}
\end{equation}
The matrices $T_i$ are required to form an affine representation of the Lie algebra of the scalar generators $Q_i$ and thus satisfy $[T_i,T_j]=\im f^k_{ij}T_k$. The $n\times n$ matrices $t_i$ span the linear part of this representation and satisfy the same commutation relation. The commutation relations between $P_\m$, $K_{\m A}$, $Q_A$ and $Q_{AB}$ then take a particularly elegant form,
\begin{align}
\notag
[P_\m,K_{\n A}]&=\im g_{\m\n}Q_A,\\
\notag
[K_{\m A},K_{\n B}]&=\im(g_{AB}J_{\m\n}+g_{\m\n}Q_{AB}),\\
\notag
[K_{\m A},Q_B]&=-\im g_{AB}P_\m,\\
\label{comm1}
[K_{\m C},Q_{AB}]&=\im(g_{AC}K_{\m B}-g_{BC}K_{\m A}),\\
\notag
[Q_A,Q_B]&=0,\\
\notag
[Q_{AB},Q_C]&=\im(g_{BC}Q_A-g_{AC}Q_B),\\
\notag
[Q_{AB},Q_{CD}]&=\im(g_{AD}Q_{BC}+g_{BC}Q_{AD}-g_{AC}Q_{BD}-g_{BD}Q_{AC}),
\end{align}
and are completely fixed by the symmetric matrix $g_{AB}$, required to be an invariant tensor under the representation $t_i$ of the algebra of $Q_i$s. Note that the first two lines of eq.~\eqref{comm1} give a precise definition of the $Q_A$s and $Q_{AB}$s. The only commutators that remain to be determined are those of $Q_i$ with $K_{\m A}$ and with the $Q_A,Q_{AB}$ subsets of scalars. These are given in terms of the objects introduced above as
\begin{equation}
\begin{split}
[Q_i,K_{\m A}]&=(t_i)^B_{\phantom BA}K_{\m B}-\im d_{Ai}P_\m,\\
[Q_i,L_{AB}]&=(T^T_iL+LT_i)_{AB}.
\end{split}
\label{comm2}
\end{equation}
Of course, all the commutators displayed in eq.~\eqref{comm1} but the first two lines are contained in eq.~\eqref{comm2} due to the fact that the $Q_A$s and $Q_{AB}$s are just a subset of all the $Q_i$s. This imposes some consistency constraints on the ``metric'' $g_{AB}$: writing $Q_A\equiv a^i_AQ_i$, $g_{AB}$ is completely fixed by the coefficients $a^i_A$ and $d_{Ai}$,
\begin{equation}
g_{AB}=-a^i_Ad_{Bi}=-a^i_Bd_{Ai}.
\label{gAB}
\end{equation}
The true content of eq.~\eqref{comm1} is thus that in order for the addition of the vector generators $K_{\m A}$ to be consistent, the algebra of scalar generators must possess a subalgebra spanned on $Q_A$ and $Q_{AB}$ with very special properties. Altogether, the Lie algebra of all the generators, $J_{\m\n}$, $P_\m$, $K_{\m A}$, $Q_i$, is uniquely determined by the subalgebra of scalar generators $Q_i$, defined by the structure constants $f^k_{ij}$, its affine representation $T_i$, and the corresponding rank-2 symmetric invariant tensor $g_{AB}$.

Let us append several remarks. First, in a typical EFT application, the scalar generators $Q_i$ will correspond to some internal, coordinate-independent symmetry. Some of them will be spontaneously broken: these are the generators responsible for the physical NG degrees of freedom of the EFT. By the same token, the vector generators $K_{\m A}$ will typically correspond to redundant, coordinate-dependent symmetries, responsible for the enhancement of scattering amplitudes in the soft limit.

There are two important exceptions to these rules of thumb. First, in order to describe physical systems with invariance under time translations, the Lie algebra must possess the corresponding generator: the Hamiltonian $H$. This can be included among the $Q_i$s; we need not specify its commutators separately. By means of eqs.~\eqref{QKcomm} and~\eqref{adler}, the Hamiltonian commutes with both $J_{\m\n}$ and $P_\m$ as expected. In principle, it need not commute with the other scalar generators $Q_i$ though. Second, our symmetry algebra is built upon the ``static'' kinematic algebra, containing translations and rotations but no boosts~\cite{FigueroaOFarrill2017a,FigueroaOFarrill2018a}. (The same Lie algebra is sometimes referred to as Aristotelian~\cite{Grosvenor2018a}.) It is, however, compatible with other kinematical algebras as well provided that we include the vector of boost generators among the $K_{\m A}$s, in particular the Poincar\'e, Galilei (or Bargmann), and Carroll algebra. It can therefore serve as a starting point for discussion of EFTs on a variety of spacetimes, relativistic or nonrelativistic alike.

Next, the most general Lie algebra of symmetry generators as presented above was derived in ref.~\cite{Bogers2018a} assuming four-dimensional Minkowski spacetime, and thus, upon replacing the Minkowski metric with the Euclidean one, applies to four spatial dimensions in the present context. The Lie algebra as shown is consistent in any number of spatial dimensions $d$, yet other contributions to the commutators, proportional to the Levi-Civita tensor, may in principle exist for $d<4$. In appendix~\ref{app:3d} we show that with the additional \emph{assumption} that the set of generators $Q_A$ is linearly independent,\footnote{This requirement is in fact necessary if we want to make sure that all the generators $K_{\m A}$, being themselves linearly independent by definition, are redundant and thus the associated would-be NG fields can be consistently eliminated from the EFT. At the present stage where we address the purely mathematical question about possible extensions of the Euclidean algebra, it is nevertheless an assumption.} no such contributions exist in the physically important case of $d=3$.

Finally, note that we have assumed that the commutation relations among the various generators do not contain any central charges, that is, any central charges that could not be absorbed into a redefinition of the generators or included among the $Q_i$s. This assumption is justified for any $d>2$, as explained in detail in Appendix~\ref{app:central}. 


\section{Galileon-like theories}
\label{sec:galileon}

It is not immediately obvious which of the Lie algebras included in the general class presented in the previous section give rise to a perturbatively well-defined EFT, by which we mean the existence of a kinetic term for all dynamical degrees of freedom and generally also some interaction terms in the Lagrangian. Instead of trying to answer this question in full generality, we explicitly demonstrate the existence of physically interesting solutions.

In this section, we will assume that $d_{Ai}=0$~\cite{Bogers2018a,Bogers2018b}. By eq.~\eqref{gAB}, this implies $g_{AB}=0$. As a result, the generators $Q_i$ and $K_{\mu A}$ span a closed subalgebra, and the affine representation $T_i$ reduces to the linear representation $t_i$. The list of nonzero commutators reads
\begin{align}
\notag
[J_{\m\n},J_{\k\l}]&=\im(g_{\m\l}J_{\n\k}+g_{\n\k}J_{\m\l}-g_{\m\k}J_{\n\l}
-g_{\n\l}J_{\m\k}),\\
\notag
[J_{\m\n},P_\l]&=\im(g_{\n\l}P_\m-g_{\m\l}P_\n),\\
\notag
[J_{\m\n},K_{\l A}]&=\im(g_{\n\l}K_{\m A}-g_{\m\l}K_{\n A}),\\
\notag
[P_\m,K_{\n A}]&=\im g_{\m\n}Q_A,\\
\label{LieGalileon}
[K_{\m A},K_{\n B}]&=\im g_{\m\n}Q_{AB},\\
\notag
[Q_i,K_{\m A}]&=(t_i)^B_{\phantom BA}K_{\m B},\\
\notag
[Q_i,Q_j]&=\im f^k_{ij}Q_k,\\
\notag
[Q_i,Q_A]&=(t_i)^B_{\phantom BA}Q_B,\\
\notag
[Q_i,Q_{AB}]&=(t_i)^C_{\phantom CA}Q_{CB}+(t_i)^C_{\phantom CB}Q_{AC}.
\end{align}
We will refer to this class of Lie algebras as ``Galileon-like.'' They are specified uniquely by the Lie algebra of scalar generators $Q_i$ and its Abelian ideal, generated by the $Q_A$s and $Q_{AB}$s. This class includes theories based on the static or Carroll algebra, but not on the Poincar\'e or Galilei (or Bargmann) algebra, where the commutator of the Hamiltonian with the boost vector receives a contribution from spatial momentum.

We have reached the point where we have to decide what physics we want to describe using our EFT. In order to be as concrete as possible at the cost of limiting the generality of our discussion, we will focus on EFTs for scalar NG modes of spontaneously broken \emph{internal} symmetry. We therefore assume, first of all, that spacetime translations and spatial rotations remain unbroken. By the same token, any of the vectors $K_{\m A}$ must either stay unbroken --- especially when it represents boosts --- or be spontaneously broken but redundant. The latter possibility requires that there is a corresponding scalar generator $Q_A$ that is spontaneously broken. This ensures that the would-be NG mode associated with $K_{\m A}$ decouples from the low-energy spectrum~\cite{Endlich2014a,Brauner2014a}. In practice, it can then be either integrated out, or more conveniently, eliminated from the EFT by imposing a suitable ``inverse Higgs constraint'' (IHC)~\cite{Ivanov1975a}.

In addition, we will assume that the scalar generators can be split into disjoint subsets,
\begin{equation}
\{Q_i\}=\{\tilde Q_i\}\sqcup\{Q_A\}\sqcup\{Q_{AB}\},
\label{disjoint}
\end{equation}
such that the $\tilde Q_i$s span a closed subalgebra, $\gal$. The algebra of scalar generators then has the structure of a semidirect product, whereby the subalgebra $\gal$ acts upon the Abelian ideals spanned on $Q_A$ and $Q_{AB}$ via the representation $t_i$. The requirement~\eqref{disjoint} includes in particular the assumptions that the algebra $\gal$ does not have any central charges and that the generators $Q_{AB}$, if nonvanishing, are linearly independent of the $Q_A$s. It was argued in ref.~\cite{Bogers2018a} that giving up either of these two assumptions leads to ill-defined EFTs.

Finally, we will assume for the sake of simplicity that the Hamiltonian commutes with all the other scalar generators, which implies by eq.~\eqref{LieGalileon} that it also commutes with $K_{\m A}$, and thus belongs to the center of the whole symmetry algebra. This assumption is natural for the static (Aristotelian) algebra, but is also consistent with the Carroll algebra, where the Hamiltonian belongs to the $Q_A$s.


\subsection{Coset construction}

With all the simplifying assumptions in place, we now set out to construct a class of Galileon-like EFTs. We employ for that purpose the version of the technique of nonlinear realizations --- oftentimes referred to as the coset construction --- designed for spacetime symmetries~\cite{Volkov1973a,Ogievetsky1974a}. We use the following parametrization for the coset space defined by the nonlinearly realized symmetries, including spacetime translations,
\begin{equation}
U(t,x,\t,\x)\equiv e^{\im tH}e^{\im x^\m P_\m}e^{\im\t^A Q_A}e^{\frac\im2\t^{AB}Q_{AB}}e^{\im\x^{\m A}K_{\m A}}e^{\im\t^a\tilde Q_a}.
\label{cosetGal}
\end{equation}
Here those of the generators $Q_A$, $Q_{AB}$, $K_{\m A}$ that are spontaneously broken enter with the corresponding NG fields. Those of the $\tilde Q_i$ generators that are spontaneously broken are labeled $\tilde Q_{a,b,\dotsc}$; those that remain unbroken will be referred to as $\tilde Q_{\a,\b,\dotsc}$ below~\cite{Andersen2014a}. The next step is to evaluate the Lie-algebra-valued Maurer-Cartan (MC) form, $\om\equiv-\im U^{-1}\dd U$. Decomposing the MC form in the basis of generators,
\begin{equation}
\om\equiv\om_HH+\om_P^\m P_\m+\om_Q^AQ_A+\frac12\om_Q^{AB}Q_{AB}+\om_K^{\m A}K_{\m A}+\Om^i\tilde Q_i,
\end{equation}
it is a matter of a straightforward exercise using the commutation relations~\eqref{LieGalileon} to determine the individual components,
\begin{align}
\notag
\om_H&=\dd t,\\
\notag
\om_P^\m&=\dd x^\m,\\
\label{MCGal}
\om^A_Q&=(e^{-\im\t^at_a})^A_{\phantom AB}(\dd\t^B-\x^B_\m\dd x^\m),\\
\notag
\om^{AB}_Q&=(e^{-\im\t^at_a})^A_{\phantom AC}(e^{-\im\t^bt_b})^B_{\phantom BD}\bigl[\dd\t^{CD}+\tfrac12\bigl(\x^C_\m\dd\x^{\m D}-\x^D_\m\dd\x^{\m C}\bigr)\bigr],\\
\notag
\om^{\m A}_K&=(e^{-\im\t^at_a})^A_{\phantom AB}\dd\xi^{\m B}.
\end{align}
The component $\Om^i$ strongly depends on the Lie algebra of $\tilde Q_i$ and cannot in general be evaluated in a closed form; it can however be calculated order by order in a series expansion in the NG fields $\t^a$~\cite{Andersen2014a}.

The first two lines of eq.~\eqref{MCGal} indicate that the geometry of the spacetime part of the coset space is not affected by the presence of the NG modes: $(\dd t,\dd x^\mu)$ together define a global basis of 1-forms, covariant under all the symmetries of the system. The fact that the MC form does not have any component proportional to $J_{\m\n}$ also shows that the spin connection is trivial. For future reference, we put down the expressions for the exterior derivative of the MC form, known as the MC structure equations,
\begin{align}
\notag
\dd\om_H=\dd\om_P^\m&=0,\\
\notag
\dd\om_Q^A&=-\im\Om^A_{\phantom AB}\w\om^B_Q+\om_P^\m\w\om^A_{K\m},\\
\label{MCeqGal}
\dd\om_Q^{AB}&=-\im\Om^A_{\phantom AC}\w\om_Q^{CB}-\im\Om^B_{\phantom BC}\w\om_Q^{AC}+\om_{K\m}^A\w\om_K^{\m B},\\
\notag
\dd\om_K^{\m A}&=-\im\Om^A_{\phantom AB}\w\om_K^{\m B},\\
\notag
\dd\Om^i&=\tfrac12f^i_{jk}\Om^j\w\Om^k,
\end{align}
where we introduced the shorthand notation $\Om^A_{\phantom AB}\equiv(t_i)^A_{\phantom AB}\Om^i$. These expressions descend directly from the commutation relations~\eqref{LieGalileon}, but can also easily be verified using the explicit expressions~\eqref{MCGal} for the MC form.

We are now finally in the position to clarify what symmetry the Lie algebra~\eqref{LieGalileon} actually describes. Within the coset construction, symmetry transformations are defined by acting from the left on the coset representative~\eqref{cosetGal} with an element of the symmetry group. By multiplying \eqref{cosetGal} in turn with $e^{\im a_0H}$, $e^{\im a^\m P_\m}$, $e^{\im\eps^AQ_A}$ and $e^{\frac\im2\eps^{AB}Q_{AB}}$, we see at once that the transformations generated by $H$, $P_\m$, $Q_A$ and $Q_{AB}$ act as trivial shifts on, respectively, $t$, $x^\m$, $\t^A$ and $\t^{AB}$. By multiplying with $e^{\im\eps^i\tilde Q_i}$, we find likewise that the Lie algebra $\gal$ leaves the spacetime coordinates intact and transforms $\t^A$, $\t^{AB}$ and $\x^{\m A}$ linearly,
\begin{equation}
\t^A\to\bigl(e^{\im\eps^it_i}\bigr)^A_{\phantom AB}\t^B,\qquad
\t^{AB}\to\bigl(e^{\im\eps^it_i}\bigr)^A_{\phantom AC}\bigl(e^{\im\eps^jt_j}\bigr)^B_{\phantom BD}\t^{CD},\qquad
\x^{\m A}\to\bigl(e^{\im\eps^it_i}\bigr)^A_{\phantom AB}\x^{\m B}.
\end{equation}
The action on the NG fields $\t^a$ is nonlinear and is described by the usual coset construction for internal symmetries~\cite{Coleman1969a,Callan1969a}. Finally, multiplying eq.~\eqref{cosetGal} from the left by $e^{\im\b^{\m A}K_{\m A}}$, we find the symmetry induced by the new vector generators $K_{\m A}$,
\begin{equation}
\t^A\to\t^A+\b^A_\m x^\m,\qquad
\t^{AB}\to\t^{AB}-\tfrac12(\b^A_\m\x^{\m B}-\b^B_\m\x^{\m A}),\qquad
\x^{\m A}\to\x^{\m A}+\b^{\m A}.
\label{symGal}
\end{equation}
The first and last of these transformation rules correspond to a trivial multi-flavor generalization of the usual Galileon symmetry~\cite{Nicolis2009a}. The transformation rule for $\t^{AB}$, along with the expression for $\om^{AB}_Q$ in eq.~\eqref{MCGal}, leads to a coupling of the $\t^A$ and $\t^{AB}$ modes, resulting in a class of theories dubbed ``twisted Galileon'' in ref.~\cite{Bogers2018a}.


\subsection{Invariant actions}

Within the coset construction, a linear representation of the unbroken subgroup is promoted to a nonlinear realization of the whole symmetry group whereby the parameters of the unbroken subgroup transformations acquire nontrivial dependence on the NG fields. (In mathematics, this is usually referred to as an induced representation of the full group.) In this way, operators invariant under the whole symmetry group can be constructed by imposing mere invariance under the unbroken subgroup. The set of components of the MC form splits up into multiplets with respect to the unbroken subgroup. The components of the MC form corresponding to broken generators transform in some linear representation of the unbroken subgroup, whereas the components corresponding to unbroken generators transform as a gauge field of the unbroken subgroup. As a consequence, the former constitute covariant building blocks for the construction of invariant actions, whereas the latter can be used to construct covariant derivatives. It is obviously very important to distinguish broken and unbroken generators at this stage.

Until now  we kept open the possibility that not all of the $Q_A$s and $K_{\m A}$s are spontaneously broken. It is however clear from eq.~\eqref{MCGal} that those which are not broken, so that there are no $\t^A$ and $\x^{\m A}$ fields, give a vanishing component of the MC form. The number of linearly independent $\om_Q^A$ components of the MC form thus equals the number of $Q_A$s that are spontaneously broken. These can be used to eliminate the redundant fields $\x^{\m A}$ by imposing a set of IHCs, which amounts to setting to zero the \emph{spatial} part of $\om_Q^A$. This is equivalent to
\begin{equation}
\x_\m^A=\de_\m\t^A.
\label{IHCGal}
\end{equation}
The rest of the last three lines of eq.~\eqref{MCGal} then gives covariant derivatives of the NG fields,
\begin{align}
\notag
\nabla_0\t^A&\equiv(e^{-\im\t^at_a})^A_{\phantom AB}\de_0\t^B,\\
\label{covderGal}
\nabla_M\t^{AB}&\equiv(e^{-\im\t^at_a})^A_{\phantom AC}(e^{-\im\t^bt_b})^B_{\phantom BD}\bigl[\de_M\t^{CD}+\tfrac12\bigl(\x^C_\m\de_M\x^{\m D}-\x^D_\m\de_M\x^{\m C}\bigr)\bigr],\\
\notag
\nabla_M\x^{\m A}&\equiv(e^{-\im\t^at_a})^A_{\phantom AB}\de_M\xi^{\m B}.
\end{align}
The covariant derivative of $\t^a$ is likewise defined using the $\Om^a$ component of the MC form, $\Om^a\equiv\nabla_M\t^a\dd x^M$.

All in all, invariant Lagrangians can now be constructed from the covariant derivatives listed in eq.~\eqref{covderGal} and \emph{their} covariant derivatives. Free indices on the covariant derivatives are to be contracted in a way preserving all linearly realized symmetries, notably spatial rotations. At the end of the day, the classification of all possible invariant Lagrangians reduces to the classification of invariant tensors of the unbroken subgroup~\cite{Andersen2014a}. Denoting in particular the unbroken part of $\gal$ as $\hal$, the nontrivial part of the task to classify invariant Lagrangians reduces to the problem of finding invariant tensors of $\hal$, which can be dealt with using standard tensor methods.

Note that the covariant derivatives listed in eq.~\eqref{covderGal} are not just covariant, but in fact manifestly invariant  under the symmetries generated by $Q_A$, $Q_{AB}$ and $K_{\m A}$. This manifest invariance has a price though. Namely, upon using the IHC~\eqref{IHCGal}, we find that there are no contributions to the MC form containing one spatial derivative per $\t^A$ (and no temporal derivatives). It looks like we cannot even construct a kinetic term for the Galileon fields $\t^A$! The resolution of this apparent paradox is by now well-known. Invariant actions can also be obtained from Lagrangians that change upon a symmetry transformation by a surface term. Such Lagrangians are sometimes referred to as quasi-invariant Lagrangians, but more often as Wess-Zumino (WZ) or Wess-Zumino-Witten terms.

The problem of finding possible WZ terms in $D$ spacetime dimensions boils down to finding invariant $(D+1)$-forms on the coset space that are closed but not exact~\cite{Witten1983a,D'Hoker1994a,D'Hoker1995b}.\footnote{There are some topological constraints that have to be satisfied for this procedure to yield a well-defined action: see refs.~\cite{Witten1983a,Davighi2018a} for details.} Closedness ensures that the $(D+1)$-form can be locally integrated to a $D$-form, which in turn defines a local Lagrangian density on the $D$-dimensional spacetime. Non-exactedness ensures that the WZ term is not equivalent to a strictly invariant Lagrangian density. In case of the original, relativistic single-flavor Galileon theory, it was shown in ref.~\cite{Goon2012a} that there are altogether $D+1$ WZ terms. In the same paper, the WZ terms were generalized to the multi-flavor case with $\gal=\gr{SO}(n)$, assuming that this is not spontaneously broken. In our previous work~\cite{Bogers2018a,Bogers2018b}, we showed that these Galileon WZ terms generalize to arbitrary $\gal$, regardless of whether it is spontaneously broken or not. There turn out to be up to $D+1$ \emph{types} of WZ terms, characterized by fully symmetric $\gal$-invariant tensors of rank $1$ to $D+1$. Just like for strictly invariant Lagrangians, the task to find possible WZ terms therefore reduces to the linear-algebraic problem of classifying invariant tensors of the corresponding symmetry group, in this case the full algebra $\gal$ (not just its unbroken part $\hal$).


\subsection{Wess-Zumino terms}
\label{subsec:WZ}

We would now like to see how to adapt the above-sketched construction of Galileon WZ terms to the present nonrelativistic Galileon-like algebra~\eqref{LieGalileon}. To that end, note that, similarly to ref.~\cite{Goon2012a}, we have at hand the total of $d+1=D$ rotationally-invariant and $\gal$-covariant ``Galileon'' $d$-forms,
\begin{equation}
\begin{split}
\gal_k^{A_1\dotsb A_k}&\equiv\eps_{\m_1\dotsb\m_d}\om_K^{\m_1 A_1}\w\dotsb\w\om_K^{\mu_k A_k}\w\om_P^{\m_{k+1}}\w\dotsb\w\om_P^{\m_d},\\
&=(d-k)!\,\om_K^{\m_1A_1}\w\dotsb\w\om_K^{\m_kA_k}\w\ho(\om_{P\m_1}\w\dotsb\w\om_{P\m_k}),\qquad
0\leq k\leq d,
\end{split}
\label{galform}
\end{equation}
where on the second line we used the Hodge star operator to simplify the index structure. For future reference, we also put down the exterior derivative of the Galileon forms, which follows immediately from eq.~\eqref{MCeqGal},
\begin{equation}
\dd\gal_k^{A_1\dotsb A_k}=-\im\sum_{j=1}^k\Om^{A_j}_{\phantom{A}B}\w\gal_k^{A_1\dotsb A_{j-1}BA_{j+1}\dotsb A_k}.
\label{dgalform}
\end{equation}
In order to get an invariant $(D+1)$-form, we now have to wedge $\gal_k^{A_1\dotsb A_k}$ into \emph{two} additional 1-forms, picked from the scalar components of the MC form.\footnote{In the relativistic case, we wedge just a single 1-form, $\om_Q^A$, into the basic Galileon $D$-forms. The fact that we can now combine two different scalar generators to form the WZ term is responsible for the larger variety of WZ terms, as compared to the relativistic case, that we are going to find below.} We have the following five schematic possibilities,
\begin{itemize}
\itemsep=0pt
\item $\om_Q^{A_1}\w\om_H\w\gal_k^{A_2\dotsb A_{k+1}}$,
\item $\Om^a\w\om_H\w\gal_k^{A_1\dotsb A_k}$,
\item $\om_Q^{A_1}\w\om_Q^{A_2}\w\gal_k^{A_3\dotsc A_{k+2}}$,
\item $\Om^a\w\om_Q^{A_1}\w\gal_k^{A_2\dotsc A_{k+1}}$,
\item $\Om^a\w\Om^b\w\gal_k^{A_1\dotsc A_k}$,
\end{itemize}
that we will discuss in turn. (There are no WZ terms that would contain $\om_Q^{AB}$, as we will explain in section~\ref{subsubsec:QAB}.) In each case, the drill is the same: we have to check for invariance and closedness. As to the invariance condition, only the transformation properties under $\gal$ need to be checked explicitly; rotational invariance is already satisfied by the proper contraction of indices in $\gal_k^{A_1\dotsb A_k}$ and invariance under $Q_A$, $Q_{AB}$ and $K_{\m A}$ is manifest, as explained above. Recall that the components of the MC form corresponding to nonlinearly realized generators transform linearly under the adjoint action of the unbroken subgroup~\cite{Coleman1969a,Callan1969a}. This implies the following transformation rules under the action of an element of $\hal$ with infinitesimal parameters $\eps^\a$,
\begin{align}
\notag
\delta\om_Q^A&=\im\eps^\a(t_\a)^A_{\phantom AB}\om_Q^B,\\
\notag
\delta\om_Q^{AB}&=\im\eps^\a(t_\a)^A_{\phantom AC}\om_Q^{CB}+\im\eps^\a(t_\a)^B_{\phantom BC}\om_Q^{AC},\\
\label{MCtransfo}
\delta\om_K^{\m A}&=\im\eps^\a(t_\a)^A_{\phantom AB}\om_K^{\m B},\\
\notag
\delta\Om^a&=-\eps^\a f^a_{\a b}\Om^b.
\end{align}
As to the closedness condition, this is most efficiently checked with the help of the MC structure equations~\eqref{MCeqGal}. If a given $(D+1)$-form passes both tests, being both invariant and closed, we have to integrate it and check whether the corresponding $D$-form potential itself is or is not invariant. If it is not, we have found a WZ term.


\subsubsection{WZ terms of type $\om_Q^{A_1}\w\om_H\w\gal_k^{A_2\dotsb A_{k+1}}$}
\label{subsubsec:typeI}

We first consider the following class of $(D+1)$-forms,
\begin{equation}
\om_{D+1}^k\equiv c_{A_1\dotsb A_{k+1}}\om_Q^{A_1}\w\om_H\w\gal_k^{A_2\dotsb A_{k+1}},\qquad
0\leq k\leq d.
\label{WZGalform1}
\end{equation}
It follows at once from eq.~\eqref{MCtransfo} that such a form is invariant if and only if
\begin{equation}
\sum_{j=1}^{k+1}(t_\a)^B_{\phantom BA_j}c_{A_1\dotsb A_{j-1}BA_{j+1}\dotsb A_{k+1}}=0,
\label{WZaux1}
\end{equation}
that is if the coupling $c_{A_1\dotsb A_{k+1}}$ is an invariant tensor of $\hal$. To check closedness, we calculate the exterior derivative using eqs.~\eqref{MCeqGal} and~\eqref{dgalform},
\begin{equation}
\begin{split}
\dd\om_{D+1}^k=&-\im\sum_{j=1}^{k+1}(t_i)^B_{\phantom BA_j}c_{A_1\dotsb A_{j-1}BA_{j+1}\dotsb A_{k+1}}\Om^i\w\om_Q^{A_1}\w\om_H\w\gal_k^{A_2\dotsb A_{k+1}}\\
&+c_{A_1\dotsb A_{k+1}}\om_H\w\om_P^\m\w\om_{K\m}^{A_1}\w\gal_k^{A_2\dotsb A_{k+1}}.
\end{split}
\label{WZaux2}
\end{equation}
The contribution of the unbroken components $\Om^\a$ of $\Om^i$ to the first line vanishes thanks to the invariance condition~\eqref{WZaux1}. The broken components $\Om^a$ must always be nonzero thanks to the presence of the NG fields $\t^a$. In order for the contributions of $\Om^a$ to eq.~\eqref{WZaux2} to vanish as well, the coupling $c_{A_1\dotsb A_{k+1}}$ must therefore be invariant under the whole algebra $\gal$, not just its unbroken part $\hal$. As to the term on the second line of eq.~\eqref{WZaux2}, this can be upon some manipulation cast as
\begin{equation}
c_{A_1\dotsb A_{k+1}}\om_H\w\om_P^\m\w\om_{K\m}^{A_1}\w\gal_k^{A_2\dotsb A_{k+1}}=\frac{k}{d-k+1}c_{A_1\dotsb A_{k+1}}\om_H\w\om_{K\m}^{A_1}\w\om_K^{\m A_2}\w\gal_{k-1}^{A_3\dotsb A_{k+1}},
\label{WZaux3}
\end{equation}
where we have used the fact that by the definition of the Galileon forms~\eqref{galform}, the coefficient $c_{A_1\dotsb A_{k+1}}$ must be symmetric in $A_2,\dotsc,A_{k+1}$. But for eq.~\eqref{WZaux3} to vanish, $c_{A_1\dotsb A_{k+1}}$ must in fact be symmetric in all its indices. We thus conclude that the $(D+1)$-form~\eqref{WZGalform1} is invariant and closed if and only if $c_{A_1\dotsb A_{k+1}}$ is a fully symmetric invariant tensor of the algebra $\gal$.

Invariance under $\gal$ ensures that $\om_{D+1}^k$ is independent of the NG fields $\t^a$, see eq.~\eqref{MCGal}. We do not have to integrate the form~\eqref{WZGalform1} in detail since, except for the factor of $\om_H$, it is identical to its relativistic counterpart. We can thus copy-paste the final result for the corresponding Lagrangian upon imposing the IHC~\eqref{IHCGal} from refs.~\cite{Bogers2018a,Bogers2018b},
\begin{equation}
\La_k=c_{A_1\dotsb A_{k+1}}\t^{A_1}G_k^{A_2\dotsb A_{k+1}},
\label{WZGalLag1}
\end{equation}
where
\begin{equation}
G_k^{A_1\dotsb A_k}\equiv\frac1{(d-k)!}\eps^{\m_1\dotsb\m_k\l_{k+1}\dotsb\l_d}\eps^{\n_1\dotsb\n_k}_{\phantom{\n_1\dotsb\n_k}\l_{k+1}\dotsb\l_d}(\de_{\m_1}\de_{\n_1}\t^{A_1})\dotsb(\de_{\m_k}\de_{\n_k}\t^{A_k}).
\end{equation}
None of the Lagrangians $\La_k$ is invariant under the Galileon symmetry~\eqref{symGal}, hence all of them constitute a genuine WZ term.

The existence of the WZ terms rests on the existence of the fully symmetric $\gal$-invariant tensor couplings $c_{A_1\dotsb A_{k+1}}$, which places a constraint on the choice of Lie algebra $\gal$ and its representation $t_i$. Real representations of many, notably all compact, Lie algebras possess a positive-definite symmetric rank-2 invariant tensor. The Lagrangian $\La_1$ then exists and gives the spatial part of the kinetic term for the Galileon fields $\t^A$. The temporal part of the kinetic term can be constructed with the help of the covariant derivative $\nabla_0\t^A$ given in eq.~\eqref{covderGal}. This ensures that there is an infinite class of perturbatively well-defined, nonrelativistic Galileon-like EFTs. All the other WZ Lagrangians $\La_k$ with $2\leq k\leq d$, as well as the further WZ terms constructed below, generate nontrivial interactions. ($\La_0$ is a tadpole and should be discarded in a perturbative EFT.)

As an aside, note that the operators $G_k^{A_1\dotsb A_k}$ descend directly from the closed forms
\begin{equation}
\tilde\gal_k^{A_1\dotsb A_k}\equiv\eps_{\m_1\dotsb\m_d}\dd\x^{\m_1A_1}\w\dotsb\w\dd\x^{\m_kA_k}\w\dd x^{\m_{k+1}}\w\dotsb\w\dd x^{\m_d}.
\label{tildegalform}
\end{equation}
These are related, but not equal, to the Galileon forms~\eqref{galform}. While $\gal_k^{A_1\dotsb A_k}$ is manifestly $\gal$-covariant but generally not closed, $\tilde\gal_k^{A_1\dotsb A_k}$ is manifestly closed but not $\gal$-covariant. The two forms coincide when $\gal$ remains fully unbroken.


\subsubsection{WZ terms of type $\Om^a\w\om_H\w\gal_k^{A_1\dotsb A_k}$}
\label{subsubsec:typeII}

Let us move to the next class of $(D+1)$-forms,
\begin{equation}
\om_{D+1}^k\equiv c_{aA_1\dotsb A_{k}}\Om^a\w\om_H\w\gal_k^{A_1\dotsb A_{k}},\qquad
0\leq k\leq d.
\label{WZGalform2}
\end{equation}
Having collected some experience in checking the invariance and closedness of the $(D+1)$-forms, we can proceed a little faster. As in the previous case, invariance requires that the coefficient $c_{aA_1\dotsb A_k}$ is an invariant tensor of $\hal$. Since it now contains two types of indices, the detailed condition looks somewhat different than eq.~\eqref{WZaux1} though,
\begin{equation}
\im f^b_{\a a}c_{bA_1\dotsb A_k}+\sum_{j=1}^{k}(t_\a)^B_{\phantom BA_j}c_{aA_1\dotsb A_{j-1}BA_{j+1}\dotsb A_{k}}=0.
\end{equation}
Let us introduce a shorthand notation for the tensor action of generators of $\gal$ on the capital indices of the coefficient,
\begin{equation}
(\Gamma_i)_{aA_1\dotsb A_k}\equiv\im\sum_{j=1}^k(t_i)^B_{\phantom BA_j}c_{aA_1\dotsb A_{j-1}BA_{j+1}\dotsb A_k}.
\label{Gammanotation}
\end{equation}
Upon working out the exterior derivative of $\om_{D+1}^k$, it turns out that the invariance and closedness conditions can be written in a compact form, respectively, as
\begin{equation}
\begin{split}
(\Gamma_\a)_{aA_1\dotsb A_k}&=f^b_{\a a}c_{bA_1\dotsb A_k},\\
(\Gamma_a)_{bA_1\dotsb A_k}-(\Gamma_b)_{aA_1\dotsb A_k}&=f^c_{ab}c_{cA_1\dotsb A_k}.
\end{split}
\label{typeIIconds}
\end{equation}
It is not obvious how to solve these conditions in full generality. It is however easy to devise a practically useful algorithm to check for the existence of their solutions case by case. In the first step, one focuses on the invariance condition. Solving this is a standard problem in group theory that amounts to finding singlets in a direct product of representations of the unbroken subalgebra $\hal$. For given $\gal$, $\hal$ and the representation $t_i$ of $\gal$ on the Galileon fields $\t^A$, this will typically yield just a few possibilities, which are then expected to be further constrained by the second condition in eq.~\eqref{typeIIconds}.

We shall now demonstrate that there is a special class of simple solutions to eq.~\eqref{typeIIconds} for which the corresponding WZ terms can be worked out explicitly. Afterwards, we will discuss possible more complex solutions.

\paragraph{Separable WZ terms.} One obvious possibility how to satisfy eq.~\eqref{typeIIconds} is to make both the left- and the right-hand side of these conditions vanish. This can be ensured if, roughly speaking, $c_{aA_1\dotsb A_k}$ behaves as an invariant tensor \emph{separately} in $a$ and in the $A_1,\dotsc,A_k$ indices. We shall call such WZ forms separable, requiring that
\begin{equation}
f^a_{ij}c_{aA_1\dotsb A_k}=(\Gamma_i)_{aA_1\dotsb A_k}=0.
\label{typeIseparable}
\end{equation}
The latter condition ensures that $c_{aA_1\dotsb A_k}$ is a fully-symmetric $\gal$-invariant tensor with respect to the indices $A_1,\dotsc,A_k$. For compact Lie algebras $\gal$, the former condition in eq.~\eqref{typeIseparable} means that $c_{aA_1\dotsb A_k}$ can only be nonzero if the index $a$ labels a generator from the center of $\gal$. In general, this condition ensures that all contributions to $\Om^a$ but $\dd\t^a$ drop out of $\om^k_{D+1}$. It is then easy to guess the corresponding $D$-form potential, $\om_D^k=c_{aA_1\dotsb A_k}\t^a\om_H\wedge\gal_k^{A_1\dotsb A_k}=c_{aA_1\dotsb A_k}\t^a\dd t\wedge\tilde\gal_k^{A_1\dotsb A_k}$, where we used the $\gal$-invariance of $c_{aA_1\dotsb A_k}$ in the capital indices. This translates, upon imposing the IHC~\eqref{IHCGal}, to the Lagrangian
\begin{equation}
\La_k=c_{aA_1\dotsb A_k}\t^aG^{A_1\dotsb A_k}_k.
\label{WZGalLag2}
\end{equation}
These Lagrangians realize trivially the Galileon symmetry~\eqref{symGal}, but nontrivially the symmetry under $\gal$. Indeed, due to the condition~\eqref{typeIseparable}, the NG field $\t^a$ inside the Lagrangian shifts under $\gal$ effectively by a mere constant and the action associated with eq.~\eqref{WZGalLag2} is invariant thanks to the fact that $G^{A_1\dotsb A_k}_k$ is a total derivative; the latter follows by inspection, or from the fact that the form~\eqref{tildegalform} is closed.

As in the previous case, eq.~\eqref{WZGalLag1}, this class of Lagrangians exists in relativistic and nonrelativistic systems alike, although this was not noticed in refs.~\cite{Bogers2018a,Bogers2018b}. The $k=0$ Lagrangian, $\La_0$, should be discarded in a perturbative EFT, being a tadpole for $\t^a$. The $k=1$ Lagrangian $\La_1$, if allowed by $\gal$-invariance, generates mixing between the Galileon and non-Galileon NG fields, $\t^A$ and $\t^a$.  Finally, the Lagrangians $\La_k$ with $k\geq2$ provide our first example of nontrivial interactions between the Galileon and non-Galileon NG fields.

\paragraph{Non-separable WZ terms.} Finding a general solution to the algebraic constraints~\eqref{typeIIconds} is beyond the scope of the present paper. We shall however outline a class of solutions that are not separable in the above-introduced sense, and thus entangle nontrivially the action of the algebra $\gal$ on the Galileon and non-Galileon fields.

To that end, note that the second of the conditions~\eqref{typeIIconds} can be satisfied identically if we find an algebra $\gal$ and its representation $t_i$ on the Galileon fields such that
\begin{equation}
(t_a)^A_{\phantom AB}=0,\qquad
f^a_{bc}=0.
\label{typeIInonseparable}
\end{equation}
The remaining first line of eq.~\eqref{typeIIconds} then requires that the tensor coupling $c_{aA_1\dotsb A_k}$ be $\hal$-invariant. Identification of WZ terms thus once again boils down to the group-theoretic problem of finding singlets in a direct product of representations of $\hal$.

In order for eq.~\eqref{typeIInonseparable} to be consistent with the fact that the matrices $t_i$ should define a representation of $\gal$, additional constraints, $f^\b_{\a a}t_\b=f^\a_{ab}t_\a=0$, must be satisfied. This can be ensured by requiring that $f^\b_{\a a}=f^\a_{ab}=0$. (This is an inevitable conclusion, should the representation $t_\a$ of the unbroken subalgebra $\hal$ be faithful.) As a consequence, the Lie algebra $\gal$ necessarily has the structure of a semidirect product,
\begin{equation}
\gal=\hal\ltimes\kal,
\label{ghk}
\end{equation}
where the subalgebra $\hal$ acts on the Abelian ideal $\kal$, spanned on the broken generators $Q_a$. In this case, the relevant components of the MC form become extremely simple,
\begin{align}
\notag
\om_Q^A&=\dd\t^A-\x^A_\m\dd x^\m,\\
\om_K^{\m A}&=\dd\x^{\m A},\\
\Om^a&=\dd\t^a.
\notag
\end{align}
The resulting WZ Lagrangian is identical to that in eq.~\eqref{WZGalLag2}, yet the constraints on the algebra $\gal$ and the coupling $c_{aA_1\dotsb A_k}$ are now quite different.

For an explicit example, take $\gal=\gr{ISO}(n)$ and $\hal=\gr{SO}(n)$, and take $t_i$ to be the vector representation of $\hal$, in which the translation generators of $\gr{ISO}(n)$ are mapped to zero. Since the Galileon fields $\t^A$ and the non-Galileon NG fields $\t^a$ now transform in the same representation of $\gr{SO}(n)$, there is an obvious solution for the $k=1$ Lagrangian,
\begin{equation}
\La_1=\d_{aA}\t^a\de_\m\de^\m\t^A.
\end{equation}
This is invariant under $\gr{SO}(n)$ but shifts by a surface term under the translation part of $\gr{ISO}(n)$. This construction can obviously be repeated for \emph{any} choice of $\hal$. The representation of $\hal$ on $\kal$, defining the semidirect product~\eqref{ghk}, can be naturally extended to the representation of the whole algebra $\gal$ on the Galileon fields $\t^A$, whereby the generators of the Abelian ideal $\kal$ are mapped to zero, as required by eq.~\eqref{typeIInonseparable}. A symmetric rank-2 $\hal$-invariant tensor $c_{aA}$ then exists for many Lie algebras and their real representations. Higher-rank symmetric invariant tensors $c_{aA_1\dotsb A_k}$ can be constructed as the $d$-tensors of $\hal$, see for instance ref.~\cite{Azcarraga1998b}.

For a reader who wonders why we resorted to the exotic-looking class of non-semisimple algebras~\eqref{ghk} with $\gal/\hal=\gr{ISO}(n)/\gr{SO}(n)$ as the prototype, we add that the superficially similar, and more natural, candidate algebra with $\gal=\gr{SO}(n+1)$ and $\hal=\gr{SO}(n)$ would not do. In this case, the Galileon fields $\t^A$ form an $(n+1)$-vector of $\gal$. The non-Galileon NG fields $\t^a$ form an $n$-vector of $\hal$, and can be conveniently encoded in a unit $(n+1)$-vector, transforming linearly under all of $\gal$, namely $\Theta^i=(\t^a,\sqrt{1-\delta_{bc}\t^b\t^c})$. A detailed analysis then shows that, at least in the $k=1$ case, the conditions~\eqref{typeIIconds} have a unique solution, the corresponding Lagrangian being proportional to
\begin{equation}
\La_1=\d_{iA}\Theta^i\de_\m\de^\m\t^A.
\end{equation}
This is, however, strictly invariant under all the symmetries and hence not a WZ term.


\subsubsection{WZ terms of type $\om_Q^{A_1}\w\om_Q^{A_2}\w\gal_k^{A_3\dotsb A_{k+2}}$}
\label{subsubsec:typeIII}

The following three classes of WZ terms are genuinely nonrelativistic. Here $\om_H$ is replaced with a scalar component of the MC form carrying a NG field, so we expect the resulting Lagrangians to depend explicitly on time derivatives of the NG fields. We start with
\begin{equation}
\om_{D+1}^k\equiv c_{A_1\dotsb A_{k+2}}\om_Q^{A_1}\w\om_Q^{A_2}\w\gal_k^{A_3\dotsb A_{k+2}},\qquad
0\leq k\leq d.
\label{WZGalform3}
\end{equation}
By the antisymmetry of the wedge product, the coefficient $c_{A_1\dotsb A_{k+2}}$ must be antisymmetric in $A_1,A_2$ and symmetric in $A_3,\dotsc,A_{k+2}$. Invariance of the form~\eqref{WZGalform3} requires $c_{A_1\dotsb A_{k+2}}$ to be an invariant tensor of $\hal$. The exterior derivative of the form can be worked out in a way very similar to eqs.~\eqref{WZaux2} and~\eqref{WZaux3} and reads
\begin{equation}
\begin{split}
\dd\om_{D+1}^k=&-\im\sum_{j=1}^{k+2}(t_i)^B_{\phantom BA_j}c_{A_1\dotsb A_{j-1}BA_{j+1}\dotsb A_{k+2}}\Om^i\w\om_Q^{A_1}\w\om_Q^{A_2}\w\gal_k^{A_3\dotsb A_{k+2}}\\
&-\frac{2k}{d-k+1}c_{A_1\dotsb A_{k+2}}\om_Q^{A_1}\w\om_{K\m}^{A_2}\w\om_K^{\m A_3}\w\gal_{k-1}^{A_4\dotsb A_{k+2}}.
\end{split}
\end{equation}
The vanishing of the first line requires $c_{A_1\dotsb A_{k+2}}$ to be an invariant tensor of the whole algebra $\gal$. The vanishing of the second line, on the other hand, indicates that it should be symmetric under the exchange of any of $A_1,A_2$ with any of $A_3,\dotsc,A_{k+2}$, which is not possible unless the latter set is empty. This time, we thus end up with a single candidate invariant and closed $(D+1)$-form,
\begin{equation}
\om^0_{D+1}=c_{AB}\om_Q^A\w\om_Q^B\w\gal_0=\eps_{\m_1\dotsb\m_d}c_{AB}\om_Q^A\w\om_Q^B\w\dd x^{\m_1}\w\dotsb\w\dd x^{\m_d},
\end{equation}
where $c_{AB}$ is an antisymmetric tensor whose full invariance under $\gal$ ensures that the form is independent of the NG fields $\t^a$. It is easy to guess the corresponding $D$-form potential,  $\om_D^0=\eps_{\m_1\dotsb\m_d}c_{AB}\t^A\dd\t^B\w\dd x^{\m_1}\w\dotsb\w\dd x^{\m_d}$. This translates to the WZ Lagrangian
\begin{equation}
\La_0=c_{AB}\t^A\de_0\t^B,
\label{WZGalLag3}
\end{equation}
which is the temporal part of the nonrelativistic (Schr\"odinger) kinetic term, making the fields $\t^A$ pairwise canonically conjugated~\cite{Nambu2004a}. As a consequence, there is one excitation branch in the spectrum associated with two degrees of freedom and its dispersion relation is as a rule quadratic~\cite{Nielsen1976a}. Such modes are nowadays called type-B NG bosons~\cite{Watanabe2011a,Watanabe2012b,Hidaka2013b}.

Interestingly, the Lagrangian~\eqref{WZGalLag3} has a much larger, accidental symmetry: it changes by a total (time) derivative upon an arbitrary spatial-coordinate-dependent shift of the fields, $\t^A\to\t^A+f^A(x)$. Such a higher-order shift symmetry is typical for derivative bilinear Lagrangians~\cite{Hinterbichler2014a,Griffin2015a}.


\subsubsection{WZ terms of type $\Om^a\w\om_Q^{A_1}\w\gal_k^{A_2\dotsb A_{k+1}}$}
\label{subsubsec:typeIV}

We move on to the next class of $(D+1)$-forms,
\begin{equation}
\om_{D+1}^k\equiv c_{aA_1\dotsb A_{k+1}}\Om^a\w\om_Q^{A_1}\w\gal_k^{A_2\dotsb A_{k+1}},\qquad
0\leq k\leq d.
\label{WZGalform4}
\end{equation}
The conditions for invariance and closedness are found in a close parallel with section~\ref{subsubsec:typeII}, we thus merely summarize the outcome of the check, using the notation~\eqref{Gammanotation}. It turns out that $c_{aA_1\dotsb A_{k+1}}$ has to be fully symmetric in the indices $A_1,\dotsc,A_{k+1}$. In addition, the following conditions must hold,
\begin{equation}
\begin{split}
(\Gamma_\a)_{aA_1\dotsb A_{k+1}}&=f^b_{\a a}c_{bA_1\dotsb A_{k+1}},\\
(\Gamma_a)_{bA_1\dotsb A_{k+1}}-(\Gamma_b)_{aA_1\dotsb A_{k+1}}&=f^c_{ab}c_{cA_1\dotsb A_{k+1}},
\end{split}
\label{typeIVconds}
\end{equation}
which are identical to eq.~\eqref{typeIIconds} upon the replacement $k\to k+1$.

Given the close analogy with section~\ref{subsubsec:typeII}, the construction of candidate invariant and closed WZ forms proceeds along the same steps; the WZ forms of the $\Om^a\w\om_Q^{A_1}\w\gal_k^{A_2\dotsb A_{k+1}}$ type are characterized by the same class of tensor couplings as the WZ forms of the $\Om^a\w\om_H\w\gal_{k+1}^{A_1\dotsb A_{k+1}}$ type. The resulting effective Lagrangians may however be quite different. Below we give some explicit examples following the division of candidate WZ forms into separable and non-separable ones as in section~\ref{subsubsec:typeII}.

\paragraph{Separable WZ terms.} These are characterized by couplings that satisfy the conditions
\begin{equation}
f^a_{ij}c_{aA_1\dotsb A_{k+1}}=(\Gamma_i)_{aA_1\dotsb A_{k+1}}=0.
\end{equation}
Similarly to the discussion of separable WZ terms in section~\ref{subsubsec:typeII}, the former condition ensures that only the $\dd\t^a$ contribution to $\Om^a$ survives in $\om_{D+1}^k$. The corresponding $D$-form potential then is
\begin{equation}
\om^k_D= c_{aA_1\dotsb A_{k+1}}\t^a\om_Q^{A_1}\w\gal_k^{A_2\dotsb A_{k+1}}=c_{aA_1\dotsb A_{k+1}}\t^a(\dd\t^{A_1}-\x^{A_1}_\m\dd x^\m)\w\tilde\gal_k^{A_2\dotsb A_{k+1}}.
\end{equation}
The fact that $\dd\om_D^k=\om_{D+1}^k$ follows from the fact that $\om_D^k$ with the $\t^a$ factor stripped off is, up to the factor $\om_H$, just the form~\eqref{WZGalform1}, which we know is closed. Upon imposing the IHC~\eqref{IHCGal}, the ensuing Lagrangian is seen to be, up to an overall factor,
\begin{equation}
\La_k=c_{aA_1\dotsb A_{k+1}}\t^a\de_0\t^{A_1}G_k^{A_2\dotsb A_{k+1}}.
\label{WZGalLag4}
\end{equation}
The Galileon symmetry~\eqref{symGal} is, as in the case of the similarly-looking Lagrangian~\eqref{WZGalLag2}, realized trivially. It is however instructive to work out a couple of examples to see that the realization of the effective shift symmetry acting on $\t^a$ can now be tricky. The first representative of the sequence of Lagrangians~\eqref{WZGalLag4} is
\begin{equation}
\La_0=c_{aA}\t^a\de_0\t^A,
\label{typeIVaux}
\end{equation}
which shifts upon $\t^a\to\t^a+\eps^a$ by a total time derivative. This is a single-derivative mixing term that, just like eq.~\eqref{WZGalLag3}, makes some of the NG fields canonically conjugated. It is interesting to see that such canonical conjugation can also mix Galileon and non-Galileon fields. The next Lagrangian in the series~\eqref{WZGalLag4} reads
\begin{equation}
\La_1=c_{aAB}\t^a\de_0\t^A\de_\m\de^\m\t^B.
\end{equation}
Here the symmetry of $c_{aAB}$ in $A,B$ is essential to ensure that upon the shift $\t^a\to\t^a+\eps^a$, the Lagrangian changes by a mere surface term,
\begin{equation}
\delta\La_1=c_{aAB}\eps^a\bigl[\de_0\bigl(\t^A\de_\m\de^\m\t^B+\tfrac12\de_\m\t^A\de^\m\t^B\bigr)-\de_\m(\t^A\de_0\de^\m\t^B)\bigr].
\end{equation}

\paragraph{Non-separable WZ terms.} Since the algebraic conditions~\eqref{typeIVconds} enforced by invariance and closedness are identical to those found in section~\ref{subsubsec:typeII}, the discussion of non-separable terms presented therein, including the constraints~\eqref{typeIInonseparable} and~\eqref{ghk}, applies here as well. In the concrete case where $\gal=\gr{ISO}(n)$ and $\hal=\gr{SO}(n)$, we find the $k=1$ Lagrangian
\begin{equation}
\La_1=\d_{aA}\t^a\de_0\t^A.
\end{equation}
This is very similar to eq.~\eqref{typeIVaux}, but was obtained under different assumptions on the symmetry algebra $\gal$ and its representation $t_i$.


\subsubsection{WZ terms of type $\Om^a\w\Om^b\w\gal_k^{A_1\dotsb A_k}$}
\label{subsubsec:typeV}

The last and most nontrivial class of $(D+1)$-forms we consider is
\begin{equation}
\om_{D+1}^k\equiv c_{abA_1\dotsb A_{k}}\Om^a\w\Om^b\w\gal_k^{A_1\dotsb A_{k}},\qquad
0\leq k\leq d.
\label{WZGalform5}
\end{equation}
The coefficient $c_{abA_1\dotsb A_k}$ is antisymmetric in $a,b$ and fully symmetric in $A_1,\dotsc,A_k$ due to the antisymmetry of the wedge product. Here we need an obvious modification of the notation~\eqref{Gammanotation},
\begin{equation}
(\Gamma_i)_{abA_1\dotsb A_k}\equiv\im\sum_{j=1}^k(t_i)^B_{\phantom BA_j}c_{abA_1\dotsb A_{j-1}BA_{j+1}\dotsb A_k}.
\end{equation}
Then for $\om_{D+1}^k$ to be invariant and closed, $c_{abA_1\dotsb A_k}$ must be an invariant tensor of $\hal$,
\begin{equation}
(\Gamma_\a)_{abA_1\dotsb A_k}=f^c_{\a a}c_{cbA_1\dotsb A_k}+f^c_{\a b}c_{acA_1\dotsb A_k}.
\label{typeVconda}
\end{equation}
In addition, the following condition must hold,
\begin{equation}
(\Gamma_c)_{abA_1\dotsb A_k}-f^d_{ab}c_{dcA_1\dotsb A_k}+\text{cyclic permutations of $a,b,c$}=0.
\label{typeVcondb}
\end{equation}
As for the WZ forms discussed in sections~\ref{subsubsec:typeII} and~\ref{subsubsec:typeIV}, it seems difficult to solve the above two conditions in general. A reasonable approach is to first focus on the $\hal$-invariance condition that can be dealt with using standard group-theoretic methods, and then impose the condition~\eqref{typeVcondb} on the (presumably few) invariant forms found. Let us work out some concrete examples.

\paragraph{Separable WZ terms.} As before, explicit solutions to the conditions~\eqref{typeVconda} and~\eqref{typeVcondb} can be found by decoupling the $a,b$ and $A_1,\dotsc,A_k$ indices. In order for the invariance and closedness conditions to be satisfied, it is thus \emph{sufficient} that $c_{abA_1\dotsb A_k}$ is a fully symmetric $\gal$-invariant tensor with respect to the $A_1,\dotsc,A_k$ indices, and that $f^a_{ij}c_{abA_1\dotsb A_k}\Om^b\w\Om^i\w\Om^j=0$ for all possible values of $A_1,\dotsc,A_k$. The latter condition then automatically guarantees that $c_{abA_1\dotsb A_k}$ is an antisymmetric $\hal$-invariant tensor in its $a,b$ indices. Now the $d$-form $\gal_k^{A_1\dotsb A_k}$, when contracted with a fully symmetric $\gal$-invariant tensor coupling, is closed. It follows that for the form~\eqref{WZGalform5} to be closed but not exact, $c_{abA_1\dotsb A_k}\Om^a\w\Om^b$ itself must be closed but not exact. For compact and connected Lie groups $G$ (stemming from the Lie algebra $\gal$), such 2-forms are classified by the second de Rham cohomology of the corresponding coset space $G/H$~\cite{D'Hoker1995b}. There is one such 2-form for every $\gr{U(1)}$ factor of $H$, or every generator of the center of its Lie algebra, $\hal$. Going back to the full $(D+1)$-form~\eqref{WZGalform5}, its $D$-form potential then is
\begin{equation}
\om_D^k=c_{\a A_1\dotsb A_k}\Om^\a\w\gal_k^{A_1\dotsb A_k},
\end{equation}
where the index $\a$ now labels generators of the center of $\hal$ and $c_{abA_1\dotsb A_k}=\frac\im2f^\a_{ab}c_{\a A_1\dotsb A_k}$. Upon imposing the IHC~\eqref{IHCGal}, we finally read off the class of WZ Lagrangians,
\begin{equation}
\La_k=c_{\a A_1\dotsb A_k}E^{M_0M_1\dotsb M_k\l_{k+1}\dotsb \l_d}\eps^{\n_1\dotsb\n_k}_{\phantom{\n_1\dotsb\n_k}\l_{k+1}\dotsb\l_d}\Om^\a_{M_0}\de_{M_1}\de_{\n_1}\t^{A_1}\dotsb\de_{M_k}\de_{\n_k}\t^{A_k},
\label{WZGalLag5}
\end{equation}
where the component $\Om^i_M$ is defined through $\Om^i\equiv\Om^i_M\dd x^M$ and we used the letter $E$ to denote the $D$-dimensional Levi-Civita symbol.

Let us have a closer look at the first two Lagrangians of the series~\eqref{WZGalLag5} to understand their content. First, we have $\La_0=c_\a\Om^\a_0$. This is a WZ term since for generators of the center of $\hal$, $\Om^\a$ transforms under $\gal$ as an Abelian gauge field, $\Om^\a\to\Om^\a+\dd\eps^\a$, where $\eps^\a$ is now a parameter of an induced $\hal$-transformation, depending on spacetime coordinates implicitly through the NG fields~\cite{Coleman1969a,Callan1969a}. This Lagrangian is responsible for the appearance of type-B NG modes in quantum many-body systems with a compact internal symmetry group, for instance in ferromagnets~\cite{Leutwyler1994a}. It is topological in nature and generates a Berry phase when the ordered ground state is driven by a time-dependent external field~\cite{Watanabe2014a,Brauner2014b}.

The next Lagrangian in the series~\eqref{WZGalLag5} reads
\begin{equation}
\La_1=c_{\a A}(\Om^\a_0\de_\m\de^\m\t^A-\Om^\a_\m\de_0\de^\m\t^A).
\label{WZaux}
\end{equation}
While it realizes trivially all symmetry transformations acting on $\t^A$, it does realize nontrivially the action of $\gal$ on $\Om^\a$. Namely, upon the Abelian gauge transformation of $\Om^\a$, the Lagrangian shifts by the surface term
\begin{equation}
\delta\La_1=c_{\a A}\bigl[\de_\m\bigl(\de_0\eps^\a\de^\m\t^A)-\de_0(\de_\m\eps^\a\de^\m\t^A\bigr)\bigr].
\end{equation}
All the Lagrangians $\La_k$ with $k\geq1$, if permitted by the required $\gal$-invariance of $c_{\a A_1\dotsb A_k}$ with respect to the $A_1,\dotsc,A_k$ indices, represent nontrivial interactions between the Galileon fields $\t^A$ and type-B NG bosons.

\paragraph{Non-separable WZ terms.} The discussion of possible non-separable WZ terms follows the same pattern as in the previous subsections. Namely, while we are not able to give a general solution to the conditions~\eqref{typeVconda} and~\eqref{typeVcondb}, the latter can be identically satisfied if eq.~\eqref{typeIInonseparable} holds, which leads us to consider non-semisimple algebras of the type~\eqref{ghk}. The $(D+1)$-form $\om_{D+1}^k$ then boils down to
\begin{equation}
\om_{D+1}^k=c_{abA_1\dotsb A_k}\dd\t^a\wedge\dd\t^b\wedge\tilde\gal_k^{A_1\dotsb A_k}.
\end{equation}
The integration of this form is trivial, leading to the class of Lagrangians
\begin{equation}
\La_k=c_{abA_1\dotsb A_k}E^{M_0M_1\dotsb M_k\l_{k+1}\dotsb \l_d}\eps^{\n_1\dotsb\n_k}_{\phantom{\n_1\dotsb\n_k}\l_{k+1}\dotsb\l_d}\t^a\de_{M_0}\t^b\de_{M_1}\de_{\n_1}\t^{A_1}\dotsb\de_{M_k}\de_{\n_k}\t^{A_k}.
\label{WZGalLag5b}
\end{equation}
This is formally similar to the Lagrangian~\eqref{WZGalLag5}, but is based on very different requirements on the Lie algebras $\gal$ and $\hal$ and the representation $t_i$ of $\gal$ on the Galileon fields. The assumed antisymmetry of $c_{abA_1\dotsb A_k}$ in $a,b$ makes the very existence of a nontrivial Lagrangian possible; whatever symmetric part of $c_{abA_1\dotsb A_k}$ might be present contributes a pure surface term to the Lagrangian~\eqref{WZGalLag5b}.

To see some concrete examples, consider $\gal/\hal=\gr{ISO}(2)/\gr{SO}(2)$. In this case, we have a natural $k=0$ Lagrangian,
\begin{equation}
\La_0=\eps_{ab}\t^a\de_0\t^b.
\end{equation}
This is invariant under $\gr{SO(2)}$ and shifts by a surface term under the spontaneously broken translation part of $\gr{ISO}(2)$. Similarly, for $\gal/\hal=\gr{ISO}(3)/\gr{SO}(3)$, we have a natural $k=1$ Lagrangian,
\begin{equation}
\La_1=\eps_{abA}\t^a(\de_0\t^b\de_\m\de^\m\t^A-\de_\m\t^b\de_0\de^\m\t^A).
\end{equation}
This is likewise manifestly invariant under $\gr{SO}(3)$, and transforms as
\begin{equation}
\delta\La_1=\eps_{abA}\eps^a\bigl[\de_\m\bigl(\de_0\t^b\de^\m\t^A\bigr)-\de_0\bigl(\de_\m\t^b\de^\m\t^A\bigr)\bigr]
\end{equation}
under the translation $\t^a\to\t^a+\eps^a$.


\subsubsection{Absence of WZ terms with $\om_Q^{AB}$}
\label{subsubsec:QAB}

With the experience accumulated by the analysis of the WZ terms above, it is now easy to understand why no WZ terms can be built using the $\om_Q^{AB}$ form. Consider the following schematic rotationally invariant $(D+1)$-form,
\begin{equation}
\om_{D+1}^k=c_{\sigma A_1\dotsb A_{k+2}}\om^\sigma\w\om_Q^{A_1A_2}\w\gal_k^{A_3\dotsb A_{k+2}},
\label{WZGalQAB}
\end{equation}
where $\om$ without a subscript can be any of $\om_H$, $\om_Q^A$, $\om_Q^{AB}$ and $\Om^a$, and $\sigma$ is the corresponding abstract index. By the antisymmetry of $\om_Q^{AB}$ in its two indices and of the wedge product, the coupling $c_{\sigma A_1\dotsb A_{k+2}}$ must be antisymmetric in $A_1,A_2$ and symmetric in $A_3,\dotsc,A_{k+2}$. As is clear from eq.~\eqref{MCeqGal}, the exterior derivative of $\om_{D+1}^k$ will contain the following term,
\begin{equation}
\dd\om_{D+1}^k\ni c_{\sigma A_1\dotsb A_{k+2}}\om^\sigma\w\om_{K\m}^{A_1}\w\om_K^{\m A_2}\w\gal_k^{A_3\dotsb A_{k+2}},
\label{omABaux}
\end{equation}
and no other contribution that would contain the same number of $\om_K^{\m A}$ 1-forms,\footnote{In the exceptional case where $\om^\sigma$ itself is of the $\om_Q^{AB}$ type, there will be two terms of the type~\eqref{omABaux}, which turn out to equal each other thanks to the symmetries of the $c_{\sigma A_1\dotsb A_{k+2}}$ tensor coupling.} hence this term has to vanish by itself, should $\om_{D+1}^k$ be closed. But this is only possible if $c_{\sigma A_1\dotsb A_{k+2}}$ is symmetric under the exchange of one of $A_1,A_2$ with one of $A_3,\dotsc,A_{k+2}$, which is forbidden by the a priori (anti)symmetry of the coupling. We conclude that indeed there are no invariant and closed forms of the type~\eqref{WZGalQAB}.


\section{DBI-like theories}
\label{sec:DBI}

Upon having investigated in detail nonrelativistic Galileon-like theories, we shall now switch gears and focus on another infinite set of EFTs, arising from the general class of Lie algebras introduced in section~\ref{sec:algebra}. In contrast to the Galileon-like theories where $g_{AB}=0$, we shall assume that $g_{AB}$ is nonsingular. The matrix $g_{AB}$ then behaves as a metric that can be used to raise and lower indices. This interpretation is supported by eq.~\eqref{comm1}, whose structure is identical to that of the pseudo-Euclidean algebra. Altogether, the generators $J_{\m\n}$, $P_\m$, $K_{\m A}$, $Q_A$ and $Q_{AB}$ span the pseudo-Euclidean algebra in a flat $(d+n)$-dimensional space equipped with the metric $g_{\m\n}\oplus g_{AB}$. The generators $Q_{AB}$ and $Q_A$ act as generators of rotations and translations in the $n$ extra dimensions, while the generators $K_{\m A}$ represent rotations between the $d$ physical dimensions and the $n$ extra dimensions. The NG fields $\t^A$ can be thought of as the fluctuations of a $d$-dimensional brane, spontaneously breaking translations in the extra dimensions~\cite{Rham2010a}.

The fact that $g_{AB}$ is invertible makes it possible to redefine the generators as~\cite{Bogers2018a,Bogers2018b}
\begin{equation}
\tilde Q_i\equiv Q_i+d_{Ai}g^{AB}Q_B.
\label{redef}
\end{equation}
It follows from eq.~\eqref{gAB} that $\tilde Q_A\equiv a^i_A\tilde Q_i=0$. At the same time, it is obvious from eq.~\eqref{redef} that \emph{any} $Q_i$ that is mapped to zero by this change of basis must be a linear combination of the $Q_A$s. The set of generators $Q_i$ thus naturally splits up into the set of $Q_A$ and those of $\tilde Q_i$ that are nonzero. We will not list all the commutation relations among the various generators upon the redefinition~\eqref{redef}. They mostly copy eqs.~\eqref{euclidean} and~\eqref{comm1} thanks to the fact that $Q_{AB}$ is unaffected by the redefinition~\eqref{redef}. We will thus focus solely on the commutators of the redefined scalars themselves, which simplify to
\begin{align}
\notag
[\tilde Q_i,K_{\m A}]&=(t_i)^B_{\phantom BA}K_{\m B},\\
\label{comm3}
[\tilde Q_i,Q_A]&=(t_i)^B_{\phantom BA}Q_B,\\
\notag
[\tilde Q_i,Q_{AB}]&=(t_i)^C_{\phantom CA}Q_{CB}+(t_i)^C_{\phantom CB}Q_{AC},\\
\notag
[\tilde Q_i,\tilde Q_j]&=\im f^k_{ij}\tilde Q_k.
\end{align}
The last line of eq.~\eqref{comm3} shows that the $\tilde Q_i$s still span a closed Lie algebra. The first three lines indicate that they act upon $K_{\m A}$, $Q_A$ and $Q_{AB}$ separately via the representation $t_i$.

Altogether, the Lie algebra of symmetry generators, denoted from now on as ``DBI-like,'' has the structure of a semidirect product. It is specified uniquely by the Lie algebra of $\tilde Q_i$, its representation $t_i$ and the invariant metric $g_{AB}$. Since the coefficients $d_{Ai}$ are now absorbed into the change of basis of the Lie algebra, the metric $g_{AB}$ can be chosen truly independently of the rest of the structure, and is only constrained by the fact that it operates in the target space of the representation $t_i$. The structure of nonrelativistic DBI-like algebras is general enough to accommodate the static (Aristotelian) kinematic algebra and the Poincar\'e algebra. The Galilei (or Bargmann) and Carroll algebras do not fit in this category, having mutually commuting boosts. For the Poincar\'e algebra, the Hamiltonian must belong to the Abelian subalgebra generated by the $Q_A$s due to the fact that it appears in the commutator of boosts with spatial translations. In case of the static algebra, we do not have a vector of boost generators, and therefore naturally expect the Hamiltonian to fit into the algebra of $\tilde Q_i$. In this case we will assume for the sake of simplicity that the Hamiltonian commutes with all the other scalar generators, which implies by eq.~\eqref{comm3} that it also commutes with all the $K_{\m A}$.


\subsection{Coset construction}

In order to work out concrete EFTs, we have to choose what physics we want to describe. Similar comments to those made in the introduction to section~\ref{sec:galileon} apply here. In particular, note that not all of the $K_{\m A}$s have to be nonlinearly realized, but those that are should be accompanied by a spontaneously broken scalar generator $Q_A$ and the corresponding NG field $\t^A$. In order to be able to employ the machinery of the coset construction, we choose the following parametrization of the coset space,
\begin{equation}
U(t,x,\t,\x)\equiv e^{\im tH}e^{\im x^\m P_\m}e^{\im\t^A Q_A}e^{\im\x^{\m A}K_{\m A}}e^{\im\t^a\tilde Q_a}.
\label{cosetDBI}
\end{equation}
This is very similar to the parametrization~\eqref{cosetGal} used for Galileon-like theories, except that the explicit $\t^{AB}$ term is missing as it is now included among the $\t^a$ fields. The fact that the $e^{\im tH}$ factor is included indicates that we have implicitly in mind the static algebra case. For the Poincar\'e algebra, this factor is to be removed and the Hamiltonian identified among the $Q_A$ generators.

The MC form of DBI-like theories takes the general form
\begin{equation}
\om\equiv\om_HH+\om_P^\m P_\m+\om_Q^AQ_A+\om_K^{\m A}K_{\m A}+\Om^i\tilde Q_i+\frac12\om_J^{\m\n}J_{\m\n}.
\end{equation}
Before we provide explicit expressions for the various components of the MC form, let us introduce some auxiliary notation. First, we define the functions
\begin{equation}
\Co(x)\equiv\cosh\sqrt x,\qquad
\Si(x)\equiv\frac{\sinh\sqrt x}{\sqrt x},
\end{equation}
which are, in spite of the appearance, both analytic functions of the argument $x$. Second, it turns out convenient to combine pairs of the redundant fields $\x^{\m A}$ by taking a product and contracting one pair of indices. This can be done in two different ways,
\begin{equation}
\Pi_\m^{\phantom\m\n}\equiv g_{AB}\x_\m^A\x^{\n B},\qquad
\Ip_A^{\phantom AB}\equiv g_{AC}\x^{\m C}\x_\m^B\equiv\x_A\cdot\x^B.
\label{PiIp}
\end{equation}
Being constructed out of the same building blocks, these matrices satisfy a set of natural intertwining relations,
\begin{equation}
[f(\Pi)]_\m^{\phantom\m\n}\x_\n^A=\x_\m^B[f(\Ip)]_B^{\phantom BA},\qquad
\x_A^\n[f(\Pi)]_\n^{\phantom\n\m}=[f(\Ip)]_A^{\phantom AB}\x_B^\m,
\end{equation}
where $f$ is an arbitrary analytic function of its argument. It is now a matter of a straightforward calculation to evaluate selected components of the MC form using the parametrization~\eqref{cosetDBI}~\cite{Bogers2018a,Bogers2018b},
\begin{align}
\notag
\om_H&=\dd t,\\
\label{MCDBI}
\om^\m_P&=\dd x^\nu(\Co\Pi)_\n^{\phantom\n\m}-\dd\t^A\xi^{\n}_A(\Si\Pi)_\n^{\phantom\n\m},\\
\notag
\om^A_Q&=(e^{-\im\t^at_a})^A_{\phantom AB}\bigl[\dd\t^C(\Co\Ip)_C^{\phantom CB}-\dd x^\mu\xi_\mu^C(\Si\Ip)_C^{\phantom CB}\bigr].
\end{align}
The components $\Om^i$ are the same as in the Galileon-like case, and only depend on the algebra of $\tilde Q_i$ and the fields $\t^a$. The component $\om_K^{\m A}$ and the spin connection $\om_J^{\m\n}$ do not seem to have a simple closed expression within the coset space parametrization used here.

Unlike in the Galileon-like case, the spacetime geometry is affected nontrivially by the presence of the NG fields. Accordingly, there is a nontrivial basis of 1-forms (vielbein) on the spacetime that must be used when extracting covariant derivatives of fields from the MC form. Following largely the notation of ref.~\cite{Brauner2014c}, we denote the covariant vielbein and its dual respectively by
\begin{equation}
e_M^N\equiv(n_M,e_M^\n),\qquad
E^M_N\equiv(V^M,E^M_\n),
\end{equation}
where $n_M$, $V^M$ are the temporal parts of the basis and $e^\n_M$, $E^M_\n$ the spatial ones. From $\om_H=\dd t$ it follows that $n_M=(1,\vek 0)$, and from $\om_P^\m$ we extract
\begin{equation}
e_0^\m=-\de_0\t^A\x_A^\n(\Si\Pi)_\n^{\phantom\n\m},\qquad
e^\m_\n=(\Co\Pi)_\n^{\phantom\n\m}-\de_\n\t^A\x_A^\l(\Si\Pi)_\l^{\phantom\l\m}.
\end{equation}
The dual vielbein is then given implicitly by
\begin{equation}
V^M=(1,-E^\m_\n e^\n_0),\qquad
E^0_\n=0,\qquad
E^\l_\n e^\m_\l=g^\m_\n.
\end{equation}

Before proceeding to the construction of actions based on the DBI-like algebras, let us clarify what symmetry these algebras generate. Working in a close parallel to the discussion of transformation rules in section~\ref{sec:galileon}, we find that the symmetries generated by $H$, $P_\m$ and $Q_A$ act as trivial shifts respectively on $t$, $x^\m$ and $\t^A$. Transformations generated by $\tilde Q_i$ act linearly thanks to the choice of coset space parametrization,
\begin{equation}
\t^A\to\bigl(e^{\im\eps^it_i}\bigr)^A_{\phantom AB}\t^B,\qquad
\x^{\m A}\to\bigl(e^{\im\eps^it_i}\bigr)^A_{\phantom AB}\x^{\m B}.
\label{Qitransfo}
\end{equation}
Finally, by multiplying eq.~\eqref{cosetDBI} from the left by $e^{\im\b^{\m A}K_{\m A}}$, and assuming now the static algebra case where the Hamiltonian necessarily commutes with all the $K_{\m A}$s, we deduce the transformation rules under the symmetry generated by $K_{\m A}$. The rules for $x^\m$ and $\t^A$ take a closed form,
\begin{equation}
x^\m\to x^\n(\Co\Pi_\b)_\n^{\phantom\n\m}+\t^A\b^{\n}_A(\Si\Pi_\b)_\n^{\phantom\n\m},\qquad
\t^A\to\t^B(\Co\Ip_\b)_B^{\phantom BA}+x^\m\b^B_\m(\Si\Ip_\b)_B^{\phantom BA},
\label{Ktransfo1}
\end{equation}
where $(\Pi_\b)_\m^{\phantom\m\n}\equiv g_{AB}\b^A_\m\b^{\n B}$ and $(\Ip_\b)_A^{\phantom AB}\equiv g_{AC}\b^{\m C}\b^B_\m$. The redundant field $\xi^{\m A}$ undergoes a constant shift to linear order,
\begin{equation}
\x^{\m A}\to\x^{\m A}+\b^{\m A}+\mathcal O(\b^2,\x^2).
\label{Ktransfo2}
\end{equation}
Finally, for those of $Q_{AB}$ that are spontaneously broken, we find
\begin{equation}
\t^{AB}\to\t^{AB}-\tfrac12(\b^A_\m\x^{\m B}-\b^{B}_\m\x^{\m A})+\mathcal O(\b^2,\t^a).
\label{Ktransfo3}
\end{equation}


\subsection{Invariant actions}

In order to be able to construct actions invariant under the above symmetry transformations, we need to identify the covariant building blocks that we can utilize for the construction. This means in particular covariant derivatives of the various fields and an invariant spacetime volume element. Covariant derivatives of NG fields are extracted from the corresponding components of the MC form with the help of the vielbein,
\begin{equation}
\om_Q^A\equiv\om_{Q,M}^A\dd x^M\equiv\nabla_N\t^Ae^N_M\dd x^M,\qquad
\Om^a\equiv\Om^a_M\dd x^M\equiv\nabla_N\t^ae^N_M\dd x^M.
\end{equation}
At the end of the day, the redundant fields $\x^{\m A}$ are eliminated by setting to zero the spatial covariant derivatives of $\t^A$. This leads to an implicit expression for $\x^{\m A}$ in terms of (ordinary) derivatives of $\t^A$,
\begin{equation}
\de_\m\t^A=\x^B_\m\left(\frac{\Si\Ip}{\Co\Ip}\right)_B^{\phantom BA}=\left(\frac{\Si\Pi}{\Co\Pi}\right)_\m^{\phantom\m\n}\x_\n^A.
\label{IHCDBI}
\end{equation}
The invariant volume element is in general given by the determinant of the vielbein. Since in our case $n_M=(1,\vek0)$, only the determinant of the spatial part of the vielbein is required. Upon some manipulation and using the IHC~\eqref{IHCDBI}, the volume element can be cast as
\begin{equation}
\dd t\,\dd^d x\,\sqrt{G},\quad\text{where}\quad
G_{\m\n}\equiv g_{\m\n}-g_{AB}\de_\m\t^A\de_\n\t^B.
\label{Gmunu}
\end{equation}
The effective metric $G_{\m\n}$ describes the extrinsic geometry of a $d$-brane embedded in a $(d+n)$-dimensional pseudo-Euclidean space. What is left of $\om_Q^A$ upon imposing the IHC~\eqref{IHCDBI} is just the temporal covariant derivative of $\t^A$,
\begin{equation}
\nabla_0\t^A=V^M\om_{Q,M}^A=(e^{-\im\t^at_a})^A_{\phantom AB}\de_0\t^C(\Co\Ip)_C^{\phantom CB}=(e^{-\im\t^at_a})^A_{\phantom AB}\de_0\t^C(\tilde G^{-1/2})_C^{\phantom CB},
\label{d0theta}
\end{equation}
where
\begin{equation}
\tilde G_A^{\phantom AB}\equiv\delta^B_A-g_{AC}\de_\m\t^C\de^\m\t^B=\delta^B_A-\de_\m\t_A\de^\m\t^B.
\label{GAB}
\end{equation}

To complete the general discussion of the construction of effective actions for DBI-like algebras, note that once the IHC~\eqref{IHCDBI} is imposed, $\om_K^{\m A}$ necessarily contains more than one derivative per each $\t^A$. At the leading order of the derivative expansion of the EFT where each of the physical NG fields carries at most one derivative, the only building blocks available for the construction of invariant Lagrangians therefore are the volume element~\eqref{Gmunu}, the temporal covariant derivative $\nabla_0\t^A$, and the covariant derivative $\nabla_M\t^a$. In a given operator, the indices of the covariant derivatives are to be contracted in a way that preserves all unbroken symmetries. The volume element~\eqref{Gmunu} gives a well-defined spatial part of the kinetic term for $\t^A$ as long as the matrix $g_{AB}$ is positive- or negative-definite. Under the same assumption, one can then construct a well-defined temporal part of the kinetic term through $g_{AB}\nabla_0\t^A\nabla_0\t^B$.\footnote{Depending on which of the generators $\tilde Q_i$ are spontaneously broken or not, this is of course not necessarily the most general invariant term with two temporal derivatives that can be added to the Lagrangian.} Hence, without any further assumptions on the choice of Lie algebra of $\tilde Q_i$ and its representation $t_i$, the DBI-like algebra is guaranteed to give a perturbatively well-defined EFT. The simplest possible theory, containing the full kinetic term for all the NG fields $\t^A$, then assumes the form
\begin{equation}
S=\int\dd t\,\dd^dx\,\sqrt G\,(1+cg_{AB}\nabla_0\t^A\nabla_0\t^B)=\int\dd t\,\dd^dx\,\sqrt G\,\bigl[1+c(\tilde G^{-1})_{AB}\de_0\t^A\de_0\t^B\bigr],
\label{DBIsimplest}
\end{equation}
where $c$ is a free parameter, and we used the invariance of $g_{AB}$ under the representation $t_i$ of the algebra of $\tilde Q_i$ to eliminate the $\t^a$ fields from the action. This is the nonrelativistic version of the multi-flavor DBI theory, discussed for instance in ref.~\cite{Cheung2017a}.


\subsection{Example: ISO(2) theory}
\label{subsec:ISO2}

Given the complexity of the expressions for the vielbein and the covariant derivatives of the NG fields, it is best to work out in detail at least one simple, concrete example of a multi-flavor DBI-like theory. Let us consider the $n=2$ case with two NG fields $\t^A$, $A=1,2$. There is a single generator $Q_{AB}\equiv\eps_{AB}Q$.  To keep things simple, we assume that there are no other scalar generators $\tilde Q_i$ except for the Hamiltonian that commutes with all the other generators. Setting $g_{AB}=\delta_{AB}$, we see from eq.~\eqref{comm1} that the scalar sector of the theory spans the $\gr{ISO}(2)$ algebra with $Q$ playing the role of rotations and $Q_A$ that of translations in the two extra dimensions. For the reader's convenience, we put together here all the relevant commutation relations of the symmetry algebra,
\begin{align}
\notag
[P_\m,K_{\n A}]&=\im g_{\m\n}Q_A,\\
\notag
[K_{\m A},K_{\n B}]&=\im(\delta_{AB}J_{\m\n}+g_{\m\n}\eps_{AB}Q),\\
\notag
[K_{\m A},Q_B]&=-\im\delta_{AB}P_\m,\\
\label{DBIalgebra}
[Q_A,Q_B]&=0,\\
\notag
[Q,Q_A]&=-\im\eps_{A}^{\phantom AB}Q_B,\\
\notag
[Q,K_{\m A}]&=-\im\eps_{A}^{\phantom AB}K_{\m B}.
\end{align}

The transformation~\eqref{Qitransfo} generated by $Q$ becomes a simple two-dimensional rotation,
\begin{equation}
\t^1\to\t^1\cos\eps-\t^2\sin\eps,\qquad
\t^2\to\t^1\sin\eps+\t^2\cos\eps.
\end{equation}
The transformation induced by $K_{\m A}$ remains the same as displayed in eqs.~\eqref{Ktransfo1} and~\eqref{Ktransfo2}. In case the generator $Q$ is also spontaneously broken, the corresponding NG field $\t$ transforms under $K_{\m A}$ by a modified version of eq.~\eqref{Ktransfo3},
\begin{equation}
\t\to\t-\tfrac12\eps_{AB}\b^A\cdot\x^B+\mathcal O(\b^2).
\end{equation}

The invariant volume element in the action is defined by the determinant of the effective metric $G_{\m\n}$~\eqref{Gmunu}, which by the Weinstein-Aronszajn identity equals the determinant of $\tilde G_{AB}$~\eqref{GAB}. (Since we chose $g_{AB}=\delta_{AB}$, there is no difference between upper and lower capital indices.) This in turn is easy to calculate explicitly in the present two-flavor case,
\begin{equation}
\tilde G=1-\delta_{AB}\de_\m\t^A\de^\m\t^B+(\de_\m\t^1)^2(\de_\n\t^2)^2-(\de_\m\t^1\de^\m\t^2)^2.
\label{volelement}
\end{equation}

It would appear that we are now ready to write down the most general effective action, at least at the leading order of the derivative expansion. But should we be careful enough, we must inspect possible Wess-Zumino terms. To that end, we first write down the MC structure equations as obtained directly from the commutation relations of our DBI-like Lie algebra,
\begin{align}
\notag
\dd\om_P^\m&=\om_Q^A\w\om_{KA}^{\m}+\om_J^{\m\n}\w\om_{P\n},\\
\dd\om_Q^A&=\eps^{A}_{\phantom AB}\om_Q\w\om_Q^B+\om_P^\m\w\om_{K\m}^A,\\
\notag
\dd\om_K^{\m A}&=\eps^A_{\phantom AB}\om_Q\w\om_K^{\m B}+\om_J^{\m\n}\w\om^A_{K\n},\\
\notag
\dd\om_Q&=\eps_{AB}\om_{K\m}^A\w\om_K^{\m B}.
\end{align}
We can scan for possible WZ terms in the same way as we did for Galileon-like WZ terms in section~\ref{subsec:WZ}, that is, by combining a set of basic rotationally invariant $d$-forms built out of $\om_P^\m$ and $\om_K^{\m A}$ with pairs of scalar MC forms picked from $\om_H$, $\om_Q^A$, $\om_Q$. There is however only one candidate $(D+1)$-form that is manifestly invariant as well as closed,\footnote{In the simplest DBI theory with a single $Q^A$ generator (denoted here as $Q$), describing the fluctuations of a flat $D$-dimensional brane embedded in a $(D+1)$-dimensional flat spacetime, there is likewise a single candidate WZ $(D+1)$-form. This is $\eps_{\mu_1\dotsb\mu_d}\om_Q\w\om_H\w\om_P^{\m_1}\w\dotsb\w\om_P^{\m_d}$, and it leads to the tadpole operator for the single NG field of the theory~\cite{Goon2012a}.}
\begin{equation}
\om_{D+1}=\eps_{\mu_1\dotsb\mu_d}\eps_{AB}\om_Q^A\w\om_Q^B\w\om_P^{\m_1}\w\dotsb\w\om_P^{\m_d}.
\label{WZDBI}
\end{equation}
This form is insensitive to whether the $\gr{SO(2)}$ subgroup generated by $Q$ is spontaneously broken or not. Indeed, if $Q$ is spontaneously broken, the corresponding NG field $\t$ enters $\om_Q^A$ through an overall $\gr{SO(2)}$ rotation, see eq.~\eqref{MCDBI}. Since $\eps_{AB}$ is an invariant tensor of $\gr{SO(2)}$, $\t$ then simply drops out of the form $\om_{D+1}$. This is a consequence of our choice of parametrization~\eqref{cosetDBI}, which ensures that the NG fields $\t^A$ transform linearly under $Q$.

Given the rather complicated expression~\eqref{MCDBI} for the MC form, it is not immediately obvious how to integrate the $(D+1)$-form~\eqref{WZDBI} and in turn directly deduce the corresponding Lagrangian. We can however take a shortcut. To the leading order in an expansion in powers of the NG fields $\t^A$, we have $\om_{D+1}=\eps_{\mu_1\dotsb\mu_d}\eps_{AB}\dd\t^A\wedge\dd\t^B\w\dd x^{\m_1}\w\dotsb\w\dd x^{\m_d}+\dotsb$. Guided by this, it is now straightforward to check that
\begin{equation}
S_\text{WZ}\equiv\int\dd t\,\dd^d x\,\eps_{AB}\t^A\de_0\t^B
\label{WZDBI2}
\end{equation}
is, in fact, exactly invariant under the whole $\gr{ISO}(2)$ algebra, and thus gives the desired, complete WZ term.

The WZ term~\eqref{WZDBI2} canonically conjugates the two NG fields, $\t^{1,2}$, leading to a single type-B NG mode in the spectrum. This is so in spite of the fact that the generators $Q_{1,2}$ of the algebra~\eqref{DBIalgebra} used to construct the action commute. The explanation is that the $\gr{ISO}(2)$ algebra becomes centrally extended upon quantization. This is a nontrivial manifestation of a phenomenon present already in the theory of a free Schr\"odinger field~\cite{Brauner2010a}. The same remark applies to some of the WZ terms constructed in section~\ref{subsec:WZ}.

Altogether, to the leading order in the derivative expansion, the effective action for our two-flavor DBI-like algebra takes the generic form
\begin{equation}
S=S_\text{WZ}+\int\dd t\,\dd^dx\,\sqrt G\,F(\nabla_0\t^A,\nabla_M\t),
\label{DBIaction}
\end{equation}
where $F$ is an arbitrary function of the indicated covariant derivatives of the NG fields. In case the $\gr{SO}(2)$ generator $Q$ is not spontaneously broken, the $\t$ field is naturally absent. In case the $\gr{SO}(2)$ \emph{is} spontaneously broken, the Lagrangian in eq.~\eqref{DBIaction} includes a term of the type $e_A\nabla_0\t^A$. The coupling $e_A$ can be physically interpreted as the vacuum expectation value $\langle Q_A\rangle$. It is easy to see that the bilinear part of such $e_A$ term is $\t\eps^{A}_{\phantom AB}e_A\de_0\t^B$. Hence this term couples $\t$ to one specific linear combination of the $\t^A$s. One can always choose a basis in the space of the $\t^A$s so that, for instance, only $e_1\neq0$ whereas $e_2=0$. Then the $e_A$ term mixes $\t$ with $\t^2$, in accord with the fact that $\langle[Q,Q_2]\rangle=\im\langle Q_1\rangle\neq0$. If both the WZ term and the $e_A$ term is present, one has to block-diagonalize the part of the Lagrangian containing a single time derivative to see the precise relation between the fields $\t^A,\t$ and the NG modes in the spectrum.


\section{Summary and conclusions}
\label{sec:summary}

In this paper, we have initiated the classification of exceptional EFTs living in spacetimes with a non-Lorentzian kinematical algebra. To that end, we first mapped the landscape of possible symmetry Lie algebras, obtained by augmenting the algebra of spacetime translations and spatial rotations with additional scalar and vector generators. The good news is that we were able to save considerable amount of effort by directly utilizing the corresponding result for Lorentz-invariant systems, obtained in ref.~\cite{Bogers2018a}.

Even better is the news that when it comes to concrete EFTs based on the thus classified Lie algebras, giving up Lorentz invariance opens the door to a new world of potentially interesting theories with nontrivially realized symmetries. We followed the basic division of EFTs into two infinite classes, dubbed ``Galileon-like'' and ``DBI-like,'' discussed previously in refs.~\cite{Bogers2018a,Bogers2018b}. Both directions turned out fruitful. For the Galileon-like algebras, we found several novel types of Wess-Zumino terms compared to Lorentz-invariant theories. These generate nontrivial interactions between Galileon and non-Galileon NG bosons, and among others open the possibility to couple Galileons to type-B NG bosons. For the DBI-like algebras, we gave a concrete example of a WZ term that turns fluctuations of a brane in two extra dimensions into a single type-B NG boson, without having to couple the system to other, non-DBI-like NG bosons. We expect many of the concrete examples of EFTs given here to share the appealing properties of relativistic Galileon and DBI theories such as the presence of scattering amplitudes with enhanced soft limit.

It should be stressed that due to the focus on the two infinite classes of EFTs based on the Galileon-like and DBI-like algebras, we did \emph{not} necessarily cover the full scope of well-defined EFTs arising from the Lie algebras classified in section~\ref{sec:algebra}. This is a challenging problem and some other possibilities were discussed previously in ref.~\cite{Bogers2018a}. To complement the analysis of multi-flavor EFTs in the main text, we show in appendix~\ref{app:1flavor} that a full classification of EFTs is possible in the simplest case of a single scalar and a single redundant vector generator. Altogether four different tentative EFTs emerge from this classification. Two of these are the one-flavor special cases of the Galileon-like and DBI-like theories discussed at length in sections~\ref{sec:galileon} and~\ref{sec:DBI}, respectively. Another is the well-known EFT for nonrelativistic superfluids, based on the Bargmann algebra. The last one is, to our best knowledge, new and represents an interesting deformation of the Galileon theory featuring a combination of the usual Galileon symmetry and temporal scaling.

Our analysis opens new questions, a detailed investigation of which would not fit in the scope of a single paper. The classification of closed invariant $(D+1)$-forms that may give rise to a WZ term, outlined in section~\ref{subsec:WZ}, suggests the possibility of novel, ``non-separable'' WZ terms where the actions of the Galileon and non-Galileon symmetries are entangled. Whether the corresponding algebraic invariance and closedness conditions can be solved in full generality remains to be seen. It would also be interesting to see whether the relativistic ``special Galileon'' symmetry~\cite{Hinterbichler2015a} has a nontrivial nonrelativistic counterpart. This could be investigated either by extending the present framework by allowing for higher-rank tensor generators~\cite{Bogers2018b}, or by checking whether a special choice of couplings for the various WZ terms listed here could further enhance the symmetry of the action~\cite{Novotny2017a}.


\acknowledgments
I would like to thank Mark Bogers for collaboration in the early stages of this project. This work has been supported by the grant no.~PR-10614 within the ToppForsk-UiS program of the University of Stavanger and the University Fund.


\appendix


\section{Symmetry algebras in $d=3$ spatial dimensions}
\label{app:3d}

As stressed in section~\ref{sec:algebra}, the general symmetry Lie algebra presented therein only includes contributions to the commutation relations that exist in an arbitrary number $d$ of spatial dimensions. Additional terms, proportional to the Levi-Civita tensor, may in principle enter the commutators for $d<4$. Here we show that in the physically most interesting case of $d=3$, such additional terms either vanish by means of the Jacobi identities, or can be absorbed into a redefinition of the generators. In order to prove this claim, we will have to make the additional assumption that the generators $Q_A$ are linearly independent, or in other words that the matrix $a^i_A$ has maximum rank.

Let us start by putting together all the commutation relations among the generators $J_{\m\n}$, $P_\m$, $K_{\m A}$, $Q_i$ in their most general form admitted by spatial rotational invariance (skipping those commutators that are known to vanish),
\begin{align}
\notag
[J_{\m\n},J_{\k\l}]&=\im(g_{\m\l}J_{\n\k}+g_{\n\k}J_{\m\l}-g_{\m\k}J_{\n\l}-g_{\n\l}J_{\m\k}),\\
\notag
[J_{\m\n},P_\l]&=\im(g_{\n\l}P_\m-g_{\m\l}P_\n),\\
\notag
[J_{\m\n},K_{\l A}]&=\im(g_{\n\l}K_{\m A}-g_{\m\l}K_{\n A}),\\
\label{Lie3d}
[P_\m,K_{\n A}]&=\im(\blue{a^i_A}g_{\m\n}Q_i+\blue{b_A}J_{\m\n}+\red{\a_A}\eps_{\m\n\l}P^\l+\red{\b_A^{\phantom AB}}\eps_{\m\n\l}K^\l_B),\\
\notag
[K_{\m A},K_{\n B}]&=\im(\blue{g_{AB}}J_{\m\n}+\blue{c^i_{AB}}g_{\m\n}Q_i+\red{\g_{AB}}\eps_{\m\n\l}P^\l+\red{\d_{AB}^{\phantom{AB}C}}\eps_{\m\n\l}K^\l_C),\\
\notag
[K_{\m A},Q_i]&=\im(\blue{d_{Ai}}P_\m+\blue{e^B_{Ai}}K_{\m B}+\red{\s_{Ai}}\eps_{\m\n\l}J^{\n\l}),\\
\notag
[Q_i,Q_j]&=\im \blue{f^k_{ij}}Q_k.
\end{align}
The a priori unknown parameters of the Lie algebra that exist in any number of spatial dimensions are highlighted in blue, following the notation introduced in ref.~\cite{Bogers2018a}. The unknown parameters that are particular to $d=3$ dimensions are highlighted in red and labeled with lowercase Greek letters. We need to find out how the unknown parameters are constrained by the various Jacobi identities that are required for the consistency of the Lie algebra. For the sake of brevity, we will refer to the Jacobi identity of the type
\begin{equation}
[A,[B,C]]+[B,[C,A]]+[C,[A,B]]=0
\end{equation}
as the ``$\{A,B,C\}$ Jacobi identity.''

We start with the contributions to the $\{P,P,K\}$ Jacobi identity, proportional to the $Q_i$ generators (referred to as the ``$Q$-terms''). The vanishing of these terms requires that
\begin{equation}
\b_A^{\phantom AB}a^i_BQ_i=\b_A^{\phantom AB}Q_B=0.
\end{equation}
It is here that we need the assumption of linear independence of $Q_A$, which immediately implies that $\b_A^{\phantom AB}=0$. Similarly, the $Q$-terms in the $\{P,K,K\}$ Jacobi identity imply, again using the linear independence of $Q_A$ and the just deduced fact that $\b_A^{\phantom AB}$ vanishes, that
\begin{equation}
\delta_{AB}^{\phantom{AB}C}=\a_A\d^C_B+\a_B\d^C_A,
\label{deltaAB}
\end{equation}
where the $\d$-symbol on the right-hand side refers to the usual Kronecker tensor. Furthermore, the $P$-terms in the $\{P,K,Q\}$ Jacobi identity allow us to solve for the $\s_{Ai}$ coefficient,
\begin{equation}
\s_{Ai}=\tfrac12e^B_{Ai}\a_B.
\label{sigmaAi}
\end{equation}
Finally, the $Q$-terms in the $\{K,K,K\}$ Jacobi identity give the constraint
\begin{equation}
(\g_{AB}Q_C-\d_{AB}^{\phantom{AB}D}c^i_{CD}Q_i)+\text{cyclic permutations of $A,B,C$}=0.
\end{equation}
Upon using the antisymmetry of the $c^i_{AB}$ coefficient in $A,B$ and the solution~\eqref{deltaAB} for $\d_{AB}^{\phantom{AB}C}$, the second term in the parentheses is seen to drop out upon summation over cyclic permutations. Hence the sum of permutations of $\g_{AB}Q_C$ alone has to vanish. Using once more the linear independence of $Q_A$ and the symmetry of $\g_{AB}$, we infer that $\g_{AB}=0$.

We are left with only three candidate contributions to the commutation relations that are specific to $d=3$ dimensions, namely the $\a_A$, $\d_{AB}^{\phantom{AB}C}$ and $\s_{Ai}$ terms. These can, however, be removed by a change of basis of the generators. Indeed, let us set
\begin{equation}
\tilde K_{\m A}\equiv K_{\m A}+\tfrac12\a_A\eps_{\m\n\l}J^{\n\l}.
\label{Kredef}
\end{equation}
This redefinition is designed so that upon using eq.~\eqref{sigmaAi}, the $\s_{Ai}$ term disappears, that is, $[\tilde K_{\m A},Q_i]=\im(d_{Ai}P_\m+e^B_{Ai}\tilde K_{\m B})$. It is straightforward to check that the $[J_{\m\n},K_{\l A}]$ commutator is unaffected by the change of basis from $K_{\m A}$ to $\tilde K_{\m A}$. Furthermore, the $\a_A$ term drops out of the commutator $[P_\m,\tilde K_{\n A}]$. Finally, it follows from eq.~\eqref{deltaAB} that
\begin{equation}
[\tilde K_{\m A},\tilde K_{\n B}]=\im(\tilde g_{AB}J_{\m\n}+c^i_{AB}g_{\m\n}Q_i),\quad\text{where}\quad
\tilde g_{AB}\equiv g_{AB}-\a_A\a_B.
\end{equation}

This completes the argument that when searching for the most general extension of the Euclidean algebra in $d=3$ spatial dimensions by adding a set of scalar and vector generators, the red-marked parameters in the commutation relations~\eqref{Lie3d} can without loss of generality be assumed to be zero. The only additional assumption that is needed for the proof is that the set of generators $Q_A\equiv a^i_AQ_i$ is linearly independent. The details of the derivation of the symmetry algebra as presented in section~\ref{sec:algebra} from the ansatz~\eqref{Lie3d} for the commutators are given in the Supplemental Material of ref.~\cite{Bogers2018a}.


\section{Absence of central charges for $d>2$}
\label{app:central}

In this appendix, we address the question to what extent the assumption of absence of central charges, made in section~\ref{sec:algebra}, limits the generality of the resulting Lie algebra structure. While several basic results on the presence or absence of central charges belong to standard textbook knowledge~\cite{Weinberg1995a}, we make the discussion self-contained.

We start by collecting all the commutators for the generators $J_{\m\n}$, $P_\m$, $K_{\m A}$, $Q_i$ as listed in section~\ref{sec:algebra}, and adding a tentative central charge to each of them (in red),
\begin{align}
\notag
[J_{\m\n},J_{\k\l}]&=\im(g_{\m\l}J_{\n\k}+g_{\n\k}J_{\m\l}-g_{\m\k}J_{\n\l}-g_{\n\l}J_{\m\k}+\red{\a^{JJ}_{\m\n\k\l}}),\\
\notag
[J_{\m\n},P_\l]&=\im(g_{\n\l}P_\m-g_{\m\l}P_\n+\red{\a^{JP}_{\m\n\l}}),\\
\notag
[J_{\m\n},K_{\l A}]&=\im(g_{\n\l}K_{\m A}-g_{\m\l}K_{\n A}+\red{\a^{JK}_{\m\n\l A}}),\\
\notag
[J_{\m\n},Q_i]&=\im\red{\a^{JQ}_{\m\n i}},\\
\label{commcentral}
[P_\m,P_\n]&=\im\red{\a^{PP}_{\m\n}},\\
\notag
[P_\m,K_{\n A}]&=\im(g_{\m\n}Q_A+\red{\a^{PK}_{\m\n A}}),\\
\notag
[P_\m,Q_i]&=\im\red{\a^{PQ}_{\m i}},\\
\notag
[K_{\m A},K_{\n B}]&=\im(g_{AB}J_{\m\n}+g_{\m\n}Q_{AB}+\red{\a^{KK}_{\m\n AB}}),\\
\notag
[K_{\m A},Q_i]&=\im[\im(t_i)^B_{\phantom BA}K_{\m B}+d_{Ai}P_\m+\red{\a^{KQ}_{\m Ai}}],\\
\notag
[Q_i,Q_j]&=\im(f^k_{ij}Q_k+\red{\a^{QQ}_{ij}}).
\end{align}
The Jacobi identities for double commutators of the generators give rise to a set of nonlinear constraints for the coefficients of the generators on the right-hand side of the commutation relations, and a set of linear constraints for the central charges. The two sets of constraints are independent from each other and thus all the constraints on the commutators discussed in section~\ref{sec:algebra} remain valid. Here we only focus on the constraints on the central charges.  We will split the discussion of the various central charges into several groups.

\paragraph{Central charges in the Euclidean algebra.} The constraint imposed by the $\{J,J,J\}$ Jacobi identity allows us to express $\a^{JJ}_{\m\n\k\l}$ in terms of a rank-2 tensor of coefficients, $\a^{JJ}_{\m\n}\equiv g^{\k\l}\a^{JJ}_{\m\k\n\l}$, as
\begin{equation}
\a^{JJ}_{\m\n\k\l}=\frac1{d-2}(g_{\m\k}\a^{JJ}_{\n\l}+g_{\n\l}\a^{JJ}_{\m\k}-g_{\m\l}\a^{JJ}_{\n\k}-g_{\n\k}\a^{JJ}_{\m\l}).
\end{equation}
The $\a^{JJ}_{\m\n\k\l}$ central charge is then seen to be trivial in that it can be removed by a redefinition of the generators,
\begin{equation}
\tilde J_{\m\n}\equiv J_{\m\n}-\frac{\a^{JJ}_{\m\n}}{d-2}.
\end{equation}
The existence of a nontrivial central charge in the commutator of rotation generators is therefore not compatible with the structure of the rotation algebra, which is a special case of a more general statement valid for all semisimple Lie algebras~\cite{Weinberg1995a}.

Assuming that the central charge $\a^{JJ}_{\m\n\k\l}$ has already been removed, we proceed to the commutator of rotation and translation generators. Here the $\{J,J,P\}$ Jacobi identity imposes a linear constraint that allows us to express $\a^{JP}_{\m\n\l}$ in terms of $\a^{JP}_\m\equiv g^{\n\l}\a^{JP}_{\m\n\l}$ as
\begin{equation}
\a^{JP}_{\m\n\l}=\frac1{d-1}(g_{\n\l}\a^{JP}_\m-g_{\m\l}\a^{JP}_\n).
\end{equation}
The central charge can then be removed from the commutator $[J_{\m\n},P_\l]$ by the redefinition
\begin{equation}
\tilde P_\m\equiv P_\m+\frac{\a^{JP}_\m}{d-1}.
\end{equation}
Note that the central charge $\a^{JK}_{\m\n\l A}$ can be dealt with in exactly the same way. We find that as a consequence of the $\{J,J,K\}$ Jacobi identity, it can always be removed by a redefinition of the $K_{\m A}$ generators.

Finally, the $\{J,P,P\}$ Jacobi identity implies that the central charge $\a^{PP}_{\m\n}$ necessarily vanishes unless $d=2$, in which case it becomes $\a^{PP}_{\m\n}=\eps_{\m\n}\a^{PP}$. This is the only possible central charge in the Euclidean algebra. It is relevant for instance for the two-dimensional dynamics of a particle in a uniform background magnetic field.

\paragraph{Central charges that only exist for $d=1$.} There can be no central charge that transforms as a vector under spatial rotations. This limits the possible existence of some of the central charges to $d=1$, in which case there are no continuous rotations. In particular, the $\{J,P,Q\}$ Jacobi identity thus excludes the $\a^{PQ}_{\m i}$ central charge unless $d=1$. The $\{P,Q,Q\}$ Jacobi identity then further constrains it by $f^k_{ij}\a^{PQ}_{\m k}=0$. Similarly, the $\{J,K,Q\}$ Jacobi identity excludes the $\a^{KQ}_{\m A i}$ unless $d=1$. In the following, we will only focus on Lie algebras in $d\geq2$ spatial dimensions, where a nontrivial Euclidean algebra to build upon exists.

\paragraph{The $\a^{JQ}$ central charge.} This is forced by the $\{J,J,Q\}$ Jacobi identity to vanish unless $d=2$, in which case it necessarily takes the form $\a^{JQ}_{\m\n i}=\eps_{\m\n}\a^{JQ}_i$. The $\{J,Q,Q\}$ Jacobi identity further constrains the coefficients $\a^{JQ}_i$ by
\begin{equation}
f^k_{ij}\a^{JQ}_k=0.
\end{equation}
Moreover, it follows from the $\{J,P,K\}$ Jacobi identity that there is no central charge in the commutator $[J_{\m\n},Q_A]$, that is, $\a^{JQ}_A\equiv a^i_A\a^{JQ}_i=0$. In a similar fashion, it follows from the $\{J,K,K\}$ Jacobi identity that there is no central charge in $[J_{\m\n},Q_{AB}]$.

\paragraph{The $\alpha^{PK}$ and $\a^{KK}$ central charges.} The $\{J,P,K\}$ Jacobi identity requires $\a^{PK}_{\m\n A}$ to take the form
\begin{equation}
\a^{PK}_{\m\n A}=g_{\m\n}\a^{PK}_A+\eps_{\m\n}\tilde\a^{PK}_A,
\end{equation}
where the first term can exist in any number of dimensions, whereas the second can only exist for $d=2$. Similarly, the $\{J,K,K\}$ Jacobi identity requires $\a^{KK}_{\m\n AB}$ to take the form
\begin{equation}
\a^{KK}_{\m\n AB}=g_{\m\n}\a^{KK}_{AB}+\eps_{\m\n}\tilde \a^{KK}_{AB},
\end{equation}
where the first term can exist in any number of dimensions, whereas the second can only exist for $d=2$.

The form of these two central charges is further constrained by the $\{P,K,Q\}$ and $\{K,K,Q\}$ Jacobi identities, which imply respectively the pairs of conditions
\begin{align}
\label{2a}
\im(t_i)^B_{\phantom BA}\a^{PK}_B+\a^{QQ}_{iA}&=0,\\
\label{2b}
\im(t_i)^B_{\phantom BA}\tilde\a^{PK}_B+d_{Ai}\a^{PP}&=0,
\end{align}
and
\begin{align}
\label{3a}
\a^{QQ}_{iAB}+\im\bigl[(t_i)^C_{\phantom CA}\a^{KK}_{CB}+(t_i)^C_{\phantom CB}\a^{KK}_{AC}\bigr]+d_{Ai}\a^{PK}_B-d_{Bi}\a^{PK}_A&=0,\\
\label{3b}
-g_{AB}\a^{JQ}_i+\im\bigl[(t_i)^C_{\phantom CA}\tilde\a^{KK}_{CB}+(t_i)^C_{\phantom CB}\tilde\a^{KK}_{AC}\bigr]+d_{Ai}\tilde\a^{PK}_B+d_{Bi}\tilde\a^{PK}_A&=0.
\end{align}
The $\a^{QQ}_{iA}$ and $\a^{QQ}_{iAB}$ coefficients arise from projecting $Q_j$ in the commutator $[Q_i,Q_j]$ to the subspace of $Q_A$ and $Q_{AB}$, respectively. A bit of straightforward algebra then shows that as a consequence of eqs.~\eqref{2a} and~\eqref{3a}, the redefinition
\begin{equation}
\tilde Q_A\equiv Q_A+\a^{PK}_A,\qquad
\tilde Q_{AB}\equiv Q_{AB}+\a^{KK}_{AB}
\end{equation}
removes the parameters $\a^{PK}_A$ and $\a^{KK}_{AB}$ from the Lie algebra. Incidentally, it also removes central charges from all commutators of $Q_A$ and $Q_{AB}$. 

\paragraph{Summary of the result.} The central charges $\a^{QQ}_{ij}$ in the scalar sector are independent of the number of spacetime dimensions. They can nevertheless always be treated as additional Abelian scalar generators, and are thus already implicitly included in the framework developed in section~\ref{sec:algebra}. Apart from these, the $\a^{PK}_A$ and $\a^{KK}_{AB}$ parameters, which we just successfully removed by a redefinition of the generators, were the only candidate central charges that could exist for $d>2$. We therefore conclude that, as asserted in the main text, our extended Euclidean algebra does not have any nontrivial central charges in $d>2$ spatial dimensions.

In $d=2$ spatial dimensions, the following commutators may feature a central charge,\footnote{For $d=2$, there can be additional contributions to the commutators between the various generators, proportional to the Levi-Civita tensor. These are not addressed in detail here.}
\begin{align}
\notag
[J_{\m\n},Q_i]&=\im\eps_{\m\n}\a^{JQ}_i,\\
\notag
[P_\m,P_\n]&=\im\eps_{\m\n}\a^{PP},\\
[P_\m,K_{\n A}]&=\im(g_{\m\n}Q_A+\eps_{\m\n}\tilde\a^{PK}_A),\\
\notag
[K_{\m A},K_{\n B}]&=\im(g_{AB}J_{\m\n}+g_{\m\n}Q_{AB}+\eps_{\m\n}\tilde\a^{KK}_{AB}),\\
\notag
[Q_i,Q_j]&=\im(f^k_{ij}Q_k+\a^{QQ}_{ij}).
\end{align}
The central charge $\a^{JQ}_i$ is constrained by the condition $f^k_{ij}\a^{JQ}_k=0$, and vanishes when projected on the subspace of the $Q_A$ or $Q_{AB}$ generators. The central charges $\a^{PP}$, $\tilde\a^{PK}_A$ and $\tilde\a^{KK}_{AB}$ are constrained by eqs.~\eqref{2b} and~\eqref{3b}. These require in particular that both $\a^{PP}$ and $\tilde\a^{PK}_A$ vanish as soon as a single component of the matrix $g_{AB}$ is nonzero.

Finally, the central charge $\a^{QQ}_{ij}$ is restricted by the generic condition
\begin{equation}
f^\ell_{ij}\a^{QQ}_{k\ell}+f^\ell_{jk}\a^{QQ}_{i\ell}+f^\ell_{ki}\a^{QQ}_{j\ell}=0.
\end{equation}
Whether or not a nontrivial solution exists depends strongly on the nature of the Lie algebra of the scalar generators $Q_i$. In any case, as stressed above, such central charges can always be included by extending the set of scalar generators appropriately.


\section{Effective theories with a single scalar}
\label{app:1flavor}

In the main text, we focused on working out two infinite classes of EFTs, based on the Galileon-like and DBI-like algebras. The price we had to pay for being able to construct concrete Lagrangians was discarding an a priori large set of potentially interesting Lie algebras that do not fall into any of these two categories. In this appendix, we demonstrate that at least in the special case of a single NG boson, corresponding to a single scalar generator and associated with it a single redundant vector generator, \emph{all} allowed Lie algebras can be found explicitly. This simple setting also makes it possible to explore the different ways to incorporate the Hamiltonian in the scalar sector of the Lie algebra.

All we have to do is to restrict the general Lie algebra structure outlined in section~\ref{sec:algebra} to a single value of the index $A$, with a single scalar generator $Q_A\equiv Q$ and a single vector generator $K_{\m A}\equiv K_\m$. We also include explicitly the Hamiltonian $H$ but no other scalar generators $Q_i$ apart from $Q$ and $H$. The full symmetry Lie algebra then consists of the Euclidean algebra of $J_{\m\n}$ and $P_\m$ augmented with the following commutation relations,
\begin{align}
\notag
[J_{\m\n},K_\l]&=\im(g_{\n\l}K_\m-g_{\m\l}K_\n),\\
\notag
[J_{\m\n},Q]&=[J_{\m\n},H]=0,\\
\notag
[P_\m,K_\n]&=\im g_{\m\n}Q,\\
\label{Lie1flavor}
[P_\m,Q]&=[P_\m,H]=0,\\
\notag
[K_\m,K_\n]&=-\im vJ_{\m\n},\\
\notag
[K_\m,Q]&=\im vP_\m,\\
\notag
[K_\m,H]&=-\im wK_\m+\im uP_\m,\\
\notag
[Q,H]&=-\im wQ.
\end{align}
The parameters $v,w$ are mutually exclusive, that is, subject to the constraint $vw=0$; this is what remains of the general condition that $g_{AB}$ be invariant under the representation of the algebra of $Q_i$ by the matrices $t_i$. As a consequence, eq.~\eqref{Lie1flavor} defines a two-parameter family of Lie algebras.


\subsection{The $w=0$ case: Galileon, DBI, and Galilei-invariant superfluid}

\begin{figure}
\includegraphics[width=\textwidth]{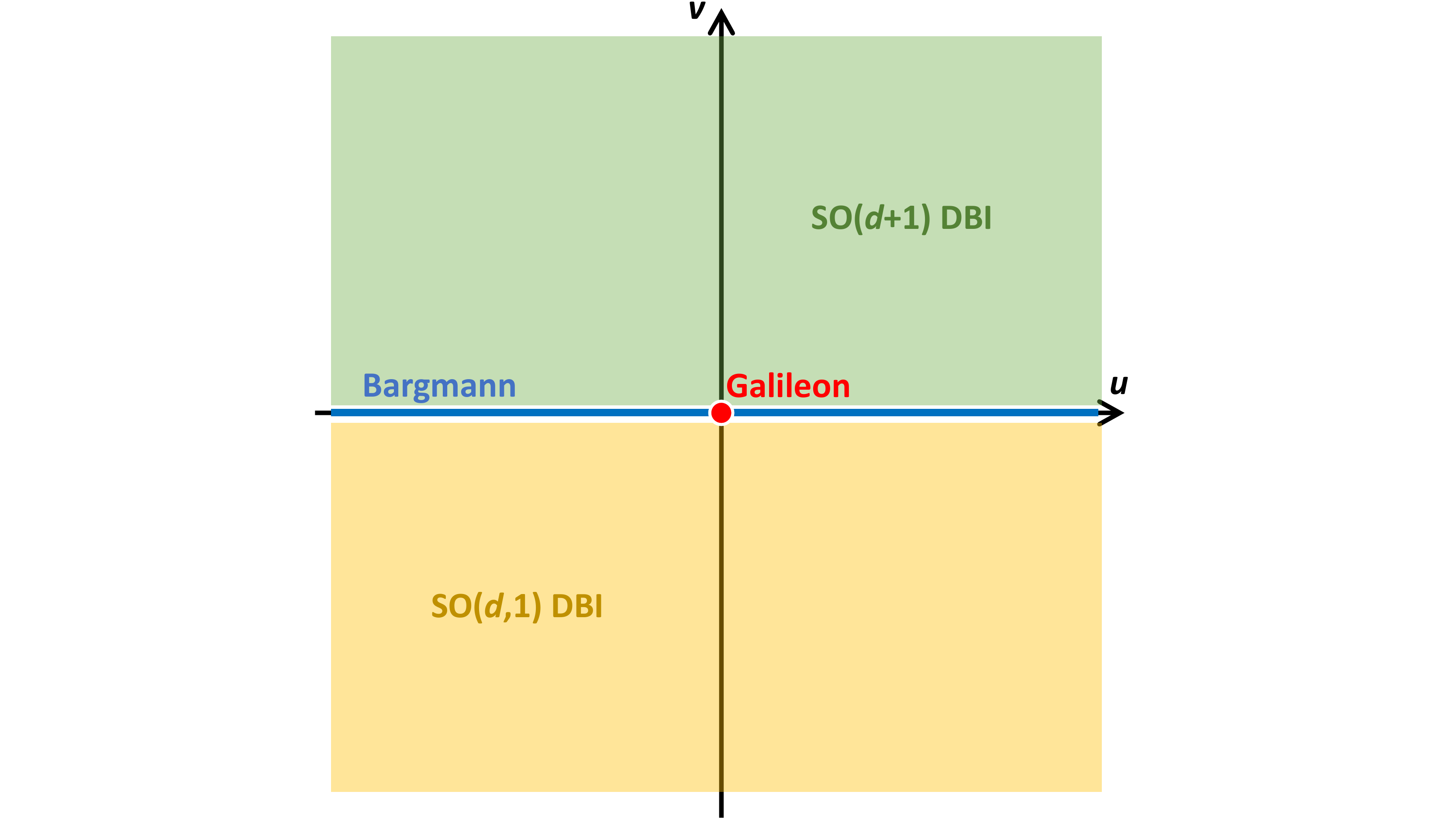}
\caption{The ``phase diagram'' of EFTs with a single scalar and a single vector generator in the plane of $u$, $v$ parameters, see eq.~\eqref{Lie1flavor}. The parameter $w$ has been set to zero. ``Bargmann'' refers to the EFT for nonrelativistic superfluids, based on the spontaneously broken Bargmann algebra.}
\label{fig:1flavor}
\end{figure}

Let us inspect what Lie algebras are included in the family~\eqref{Lie1flavor}, starting with the class of $w=0$ algebras. The three possibilities that we find and work out in detail below are presented in a graphic form in figure~\ref{fig:1flavor}.

\paragraph{The $u=v=0$ case: Galileon.} When all the parameters $u$, $v$, $w$ are zero, the only nontrivial commutators contained in eq.~\eqref{Lie1flavor} are
\begin{equation}
\begin{split}
[J_{\m\n},K_\l]&=\im(g_{\n\l}K_\m-g_{\m\l}K_\n),\\
[P_\m,K_\n]&=\im g_{\m\n}Q.
\end{split}
\label{1flavorGalileon}
\end{equation}
This is the single-flavor nonrelativistic Galileon algebra, a special case of the class of Galileon-like algebras analyzed in section~\ref{sec:galileon}. All the conclusions made therein regarding the symmetry transformations and coset construction of effective Lagrangians remain valid here. Due to the absence of any other scalar generators than $Q$ and $H$, only the WZ terms derived in section~\ref{subsubsec:typeI} are relevant for this single-flavor case. These constitute a verbatim copy of the relativistic Galileon Lagrangians, originally derived in ref.~\cite{Nicolis2009a}, upon reinterpreting the Greek indices as describing spatial coordinates only.

\paragraph{The $v\neq0$ case: DBI.} In this case, the parameter $u$ can be consistently removed from the Lie algebra by redefining the Hamiltonian, $\tilde H\equiv H-\frac uvQ$. Subsequently, the parameter $v$ itself can be reduced to $\text{sgn}\,v\equiv s$ by rescaling the $K_\m$ and $Q$ generators by $\sqrt{|v|}$. We thus end up with a discrete set of two Lie algebras, following from eq.~\eqref{Lie1flavor},
\begin{align}
\notag
[J_{\m\n},K_\l]&=\im(g_{\n\l}K_\m-g_{\m\l}K_\n),\\
[P_\m,K_\n]&=\im g_{\m\n}Q,\\
\notag
[K_\m,K_\n]&=-\im sJ_{\m\n},\\
\notag
[K_\m,Q]&=\im sP_\m.
\end{align}
This is the single-flavor special case of the class of nonrelativistic DBI-like algebras investigated in section~\ref{sec:DBI}. Namely, the scalar $Q$ acts as a generator of translations in an extra dimension, whereas the vector $K_\m$ generates rotations between the $d$ physical spatial dimensions and the extra dimension. The sign of the metric in the extra dimension is fixed by $s$. Hence, for $s=1$ the Lie algebra of symmetry generators is isomorphic to the Euclidean algebra $\gr{SO}(d+1)\ltimes\R^{d+1}$, whereas for $s=-1$ it is isomorphic to $\gr{SO}(d,1)\ltimes\R^{d+1}$. (The Hamiltonian is separate and commutes with all the other generators.)

It follows from the discussion in section~\ref{sec:DBI} that to the leading order of the derivative expansion where each NG field factor $\t$ carries one derivative, an invariant action can be built out of the invariant volume element, $\dd t\,\dd^d x\,\sqrt{1+s(\de_\m\t)^2}$, and the temporal covariant derivative of the NG field, equal to
\begin{equation}
\nabla_0\t=\frac{\de_0\t}{\sqrt{1+s(\de_\m\t)^2}}.
\end{equation}
To the leading order of the derivative expansion, the most general effective action for the single-flavor nonrelativistic DBI theory therefore assumes the form
\begin{equation}
S=\int\dd t\,\dd^dx\,\sqrt{1+s(\de_\m\t)^2}\,F(\nabla_0\t),
\end{equation}
where $F(\nabla_0\t)$ is an arbitrary, sufficiently smooth function, only constrained by the requirement that it admits a series expansion in powers of $\de_M\t$. In the formal series expansion, $F(\nabla_0\t)=\sum_{n=0}^\infty c_n(\nabla_0\t)^n$, the $c_0$ term leads to an action describing the induced spatial geometry of a $d$-dimensional flat brane, fluctuating in the extra dimension. The $c_1$ term drops out of the action, being a total time derivative. The $c_2$ term is needed alongside the $c_0$ term to get a full, both spatial and temporal, kinetic term for $\t$. All the higher-order terms $c_n$ with $n\geq3$ parametrize self-interactions of the NG mode $\t$.

\paragraph{The $v=0$, $u\neq0$ case: Galilei-invariant superfluid.} In this case, the parameter $u$ can be set to one by rescaling the Hamiltonian. We thus end up with a unique Lie algebra, whose only nontrivial commutators following from eq.~\eqref{Lie1flavor} are
\begin{align}
\notag
[J_{\m\n},K_\l]&=\im(g_{\n\l}K_\m-g_{\m\l}K_\n),\\
[P_\m,K_\n]&=\im g_{\m\n}Q,\\
\notag
[K_\m,H]&=\im P_\m.
\end{align}
This is the Bargmann algebra where $K_\m$ is the Galilei boost and $Q$ is the central charge. In quantum many-body systems, $Q$ can be interpreted as the operator of particle number. An EFT based on its spontaneous breaking then describes nonrelativistic superfluids.

The construction of EFTs for nonrelativistic superfluids based on the spacetime and internal symmetries alone is by now well-known~\cite{Greiter1989a,Son2006a}. For the reader's convenience, we will however briefly summarize the main steps. In line with the notation used in sections~\ref{sec:galileon} and~\ref{sec:DBI}, we parametrize the coset space, spanning nonlinearly realized symmetries, as
\begin{equation}
U(t,x,\t,\x)\equiv e^{\im tH}e^{\im x^\m P_\m}e^{\im\t Q}e^{\im\x^{\m}K_{\m}}.
\label{cosetBargmann}
\end{equation}
The symmetries generated by $H$, $P_\m$ and $Q$ act simply as shifts on $t$, $x^\m$ and $\t$, respectively. The only nontrivial transformation is induced by a Galilei boost with parameter $\b^\m$, under which
\begin{equation}
x^\m\to x^\m-\b^\m t,\qquad
\t\to\t+\b_\m x^\m-\tfrac12\b^2t,\qquad
\x^\m\to\x^\m+\b^\m.
\end{equation}
A few lines of calculation give the nonzero components of the MC form,
\begin{align}
\notag
\om_H&=\dd t,\\
\om_P^\m&=\dd x^\m+\x^\m\dd t,\\
\notag
\om_K^\m&=\dd\x^\m,\\
\notag
\om_Q&=\dd\t-\x_\m\dd x^\m-\tfrac12\x^2\dd t.
\end{align}
Using the same notation as in section~\ref{sec:DBI}, we find from $\om_H$ and $\om_P^\m$ that the covariant vielbein equals
\begin{equation}
n_M=(1,\vek0),\qquad
e^\n_M=(\x^\n,g^\n_\m),
\end{equation}
whereas its dual reads
\begin{equation}
V^M=(1,-\x^\m),\qquad
E^M_\n=(0,g^\m_\n).
\end{equation}
The corresponding invariant volume element is $\dd t\,\dd^d x$. The spatial part of $\om_Q$ provides the IHC necessary to eliminate the redundant field $\x^\m$; it takes the same form as for Galileon-like theories, $\x_\m=\de_\m\t$. The rest of $\om_Q$ gives the temporal covariant derivative of $\t$,
\begin{equation}
\nabla_0\t\equiv V^M\om_{Q,M}=\de_0\t-\x^\m\de_\m\t+\tfrac12\x^2\xrightarrow{\text{IHC}}\de_0\t-\tfrac12(\de_\m\t)^2.
\end{equation}
This is invariant under all the symmetries, and at the same time it is the only covariant operator that contains at most one derivative per $\t$. To the leading order of the derivative expansion, the effective Lagrangian for a nonrelativistic, Galilei-invariant superfluid is therefore given by an arbitrary function $F(\nabla_0\t)$. The form of this function can be constrained by matching to the equation of state of the superfluid~\cite{Greiter1989a,Son2006a}.


\subsection{The $w\neq0$ case: deformed Galileon}

In this case, one necessarily has $v=0$. Also, the parameter $u$ can be consistently removed by redefining $\tilde K_\m\equiv K_\m-\frac uwP_\m$. Finally, the parameter $w$ can be set to one by rescaling the Hamiltonian. At the end of the day, we thus find a unique Lie algebra, determined by the commutation relations
\begin{align}
\notag
[J_{\m\n},K_\l]&=\im(g_{\n\l}K_\m-g_{\m\l}K_\n),\\
\label{1flavorDeformedGalileon}
[P_\m,K_\n]&=\im g_{\m\n}Q,\\
\notag
[K_\m,H]&=-\im K_\m,\\
\notag
[Q,H]&=-\im Q,
\end{align}
This can be thought of as the Galileon algebra~\eqref{1flavorGalileon} deformed by temporal scaling acting on the $K_\m$ and $Q$ generators. From the group-theoretic point of view, both algebras~\eqref{1flavorGalileon} and~\eqref{1flavorDeformedGalileon} have the structure of a semidirect product of the subalgebra spanned on $\{J_{\m\n},H\}$, acting on the subalgebra spanned on $\{P_\m,K_\m,Q\}$. The action is, however, different in the two cases.

Let us see how the difference is reflected in the coset construction of invariant actions. Using the same coset space parametrization as in eq.~\eqref{cosetBargmann}, we find that the time and space translations generated by $H$ and $P_\m$ act as trivial shifts of $t$ and $x^\m$, respectively. Under a transformation generated by $Q$, with parameter $\eps$, the NG field $\t$ shifts as
\begin{equation}
\t\to\t+e^t\eps.
\end{equation}
Similarly, under a transformation generated by $K_\m$, with parameter $\b^\m$, the scalar and vector fields shift respectively as
\begin{equation}
\t\to\t+e^t\b_\m x^\m,\qquad
\x^\m\to\x^\m+e^t\b^\m.
\end{equation}
The MC form is evaluated by a simple calculation and reads
\begin{align}
\notag
\om_H&=\dd t,\\
\om_P^\m&=\dd x^\m,\\
\notag
\om_K^\m&=\dd\x^\m-\x^\m\dd t,\\
\notag
\om_Q&=\dd\t-\t\dd t-\x_\m\dd x^\m.
\end{align}
The geometry of the flat spacetime is not affected by the spontaneously broken symmetry, that is, the vielbein arising from $\om_H$ and $\om_P^\m$ is trivial. The vector field $\x^\m$ is eliminated by setting the spatial part of $\om_Q$ to zero, which corresponds to the IHC $\x_\m=\de_\m\t$. Invariant actions are then built out of the following covariant derivatives, extracted from the temporal part of $\om_Q$ and from $\om_K^\m$,
\begin{align}
\notag
\nabla_0\t&=\de_0\t-\t,\\
\nabla_0\x^\m&=\de_0\x^\m-\x^\m\xrightarrow{\text{IHC}}\de_0\de^\m\t-\de^\m\t,\\
\notag
\nabla_\n\x^\m&=\de_\n\x^\m\xrightarrow{\text{IHC}}\de_\n\de^\m\t.
\end{align}
While the deformed Galileon structure looks certainly interesting, it is not clear whether a perturbatively well-defined EFT can be constructed using the above building blocks.


\bibliographystyle{JHEP}
\bibliography{references}

\providecommand{\href}[2]{#2}\begingroup\raggedright\begin{thebibliography}{10}

\bibitem{Coleman1969a}
S.~R. Coleman, J.~Wess and B.~Zumino, \emph{{Structure of Phenomenological
  Lagrangians. I}},
  \href{http://dx.doi.org/10.1103/PhysRev.177.2239}{\emph{Phys. Rev.} {\bf 177}
  (1969) 2239--2247}.

\bibitem{Callan1969a}
C.~G. Callan, S.~Coleman, J.~Wess and B.~Zumino, \emph{{Structure of
  phenomenological Lagrangians. II}},
  \href{http://dx.doi.org/10.1103/PhysRev.177.2247}{\emph{Phys. Rev.} {\bf 177}
  (1969) 2247--2250}.

\bibitem{Leutwyler1994b}
H.~Leutwyler, \emph{On the foundations of chiral perturbation theory},
  \href{http://dx.doi.org/10.1006/aphy.1994.1094}{\emph{Ann. Phys.} {\bf 235}
  (1994) 165--203}, [\href{http://arxiv.org/abs/hep-ph/9311274}{{\tt
  hep-ph/9311274}}].

\bibitem{Leutwyler1994a}
H.~Leutwyler, \emph{{Nonrelativistic effective Lagrangians}},
  \href{http://dx.doi.org/10.1103/PhysRevD.49.3033}{\emph{Phys. Rev.} {\bf D49}
  (1994) 3033--3043}, [\href{http://arxiv.org/abs/hep-ph/9311264}{{\tt
  hep-ph/9311264}}].

\bibitem{Watanabe2014a}
H.~Watanabe and H.~Murayama, \emph{{Effective Lagrangian for Nonrelativistic
  Systems}}, \href{http://dx.doi.org/10.1103/PhysRevX.4.031057}{\emph{Phys.
  Rev.} {\bf X4} (2014) 031057}, [\href{http://arxiv.org/abs/1402.7066}{{\tt
  1402.7066}}].

\bibitem{Andersen2014a}
J.~O. Andersen, T.~Brauner, C.~P. Hofmann and A.~Vuorinen, \emph{{Effective
  Lagrangians for quantum many-body systems}},
  \href{http://dx.doi.org/10.1007/JHEP08(2014)088}{\emph{JHEP} {\bf 08} (2014)
  088}, [\href{http://arxiv.org/abs/1406.3439}{{\tt 1406.3439}}].

\bibitem{Brauner2010a}
T.~Brauner, \emph{{Spontaneous Symmetry Breaking and Nambu-Goldstone Bosons in
  Quantum Many-Body Systems}},
  \href{http://dx.doi.org/10.3390/sym2020609}{\emph{Symmetry} {\bf 2} (2010)
  609--657}, [\href{http://arxiv.org/abs/1001.5212}{{\tt 1001.5212}}].

\bibitem{Watanabe2020a}
H.~Watanabe, \emph{{Counting Rules of Nambu-Goldstone Modes}},
  \href{http://dx.doi.org/10.1146/annurev-conmatphys-031119-050644}{\emph{Ann.
  Rev. Condensed Matter Phys.} {\bf 11} (2020) 169},
  [\href{http://arxiv.org/abs/1904.00569}{{\tt 1904.00569}}].

\bibitem{Beekman2019a}
A.~J. Beekman, L.~Rademaker and J.~van Wezel, \emph{{An Introduction to
  Spontaneous Symmetry Breaking}},
  \href{http://dx.doi.org/10.21468/SciPostPhysLectNotes.11}{\emph{SciPost Phys.
  Lect. Notes} {\bf 11} (2019) }, [\href{http://arxiv.org/abs/1909.01820}{{\tt
  1909.01820}}].

\bibitem{AlvarezGaume2020a}
L.~\'Alvarez-Gaum\'e, D.~Orlando and S.~Reffert, \emph{{Selected Topics in the
  Large Quantum Number Expansion}},
  \href{http://arxiv.org/abs/2008.03308}{{\tt 2008.03308}}.

\bibitem{Ivanov1975a}
E.~Ivanov and V.~I. Ogievetsky, \emph{{The Inverse Higgs Phenomenon in
  Nonlinear Realizations}}, {\emph{Teor. Mat. Fiz.} {\bf 25} (1975) 164--177}.

\bibitem{Low2002a}
I.~Low and A.~V. Manohar, \emph{{Spontaneously Broken Spacetime Symmetries and
  Goldstone's Theorem}},
  \href{http://dx.doi.org/10.1103/PhysRevLett.88.101602}{\emph{Phys. Rev.
  Lett.} {\bf 88} (2002) 101602},
  [\href{http://arxiv.org/abs/hep-th/0110285}{{\tt hep-th/0110285}}].

\bibitem{Nambu2004a}
Y.~Nambu, \emph{{Spontaneous Breaking of Lie and Current Algebras}},
  \href{http://dx.doi.org/10.1023/B:JOSS.0000019827.74407.2d}{\emph{J. Stat.
  Phys.} {\bf 115} (2004) 7--17}.

\bibitem{Watanabe2013a}
H.~Watanabe and H.~Murayama, \emph{{Redundancies in Nambu-Goldstone Bosons}},
  \href{http://dx.doi.org/10.1103/PhysRevLett.110.181601}{\emph{Phys. Rev.
  Lett.} {\bf 110} (2013) 181601}, [\href{http://arxiv.org/abs/1302.4800}{{\tt
  1302.4800}}].

\bibitem{Brauner2014a}
T.~Brauner and H.~Watanabe, \emph{{Spontaneous breaking of spacetime symmetries
  and the inverse Higgs effect}},
  \href{http://dx.doi.org/10.1103/PhysRevD.89.085004}{\emph{Phys. Rev.} {\bf
  D89} (2014) 085004}, [\href{http://arxiv.org/abs/1401.5596}{{\tt
  1401.5596}}].

\bibitem{Brauner2020a}
T.~Brauner, \emph{{Noether currents of locally equivalent symmetries}},
  \href{http://dx.doi.org/10.1088/1402-4896/ab50a5}{\emph{Phys. Scr.} {\bf 95}
  (2020) 035004}, [\href{http://arxiv.org/abs/1910.12224}{{\tt 1910.12224}}].

\bibitem{Cheung2015a}
C.~Cheung, K.~Kampf, J.~{Novotn\'y} and J.~Trnka, \emph{{Effective Field
  Theories from Soft Limits of Scattering Amplitudes}},
  \href{http://dx.doi.org/10.1103/PhysRevLett.114.221602}{\emph{Phys. Rev.
  Lett.} {\bf 114} (2015) 221602}, [\href{http://arxiv.org/abs/1412.4095}{{\tt
  1412.4095}}].

\bibitem{Watanabe2014b}
H.~Watanabe and A.~Vishwanath, \emph{{Criterion for stability of Goldstone
  Modes and Fermi Liquid behavior in a metal with broken symmetry}},
  \href{http://dx.doi.org/10.1073/pnas.1415592111}{\emph{Proc. Nat. Acad. Sci.}
  {\bf 111} (2014) 16314--16318}, [\href{http://arxiv.org/abs/1404.3728}{{\tt
  1404.3728}}].

\bibitem{Rothstein2018a}
I.~Z. Rothstein and P.~Shrivastava, \emph{{Symmetry realization via a dynamical
  inverse Higgs mechanism}},
  \href{http://dx.doi.org/10.1007/JHEP05(2018)014}{\emph{JHEP} {\bf 05} (2018)
  014}, [\href{http://arxiv.org/abs/1712.07795}{{\tt 1712.07795}}].

\bibitem{Kampf2020a}
K.~Kampf, J.~Novotn{\'y}, M.~Shifman and J.~Trnka, \emph{{New Soft Theorems for
  Goldstone Boson Amplitudes}},
  \href{http://dx.doi.org/10.1103/PhysRevLett.124.111601}{\emph{Phys. Rev.
  Lett.} {\bf 124} (2020) 111601}, [\href{http://arxiv.org/abs/1910.04766}{{\tt
  1910.04766}}].

\bibitem{Cheung2016a}
C.~Cheung, K.~Kampf, J.~{Novotn\'y}, C.-H. Shen and J.~Trnka, \emph{{On-Shell
  Recursion Relations for Effective Field Theories}},
  \href{http://dx.doi.org/10.1103/PhysRevLett.116.041601}{\emph{Phys. Rev.
  Lett.} {\bf 116} (2016) 041601}, [\href{http://arxiv.org/abs/1509.03309}{{\tt
  1509.03309}}].

\bibitem{Cheung2017a}
C.~Cheung, K.~Kampf, J.~Novotn{\'y}, C.-H. Shen and J.~Trnka, \emph{{A periodic
  table of effective field theories}},
  \href{http://dx.doi.org/10.1007/JHEP02(2017)020}{\emph{JHEP} {\bf 02} (2017)
  020}, [\href{http://arxiv.org/abs/1611.03137}{{\tt 1611.03137}}].

\bibitem{Bogers2018a}
M.~P. Bogers and T.~Brauner, \emph{{Geometry of Multiflavor Galileon-Like
  Theories}},
  \href{http://dx.doi.org/10.1103/PhysRevLett.121.171602}{\emph{Phys. Rev.
  Lett.} {\bf 121} (2018) 171602}, [\href{http://arxiv.org/abs/1802.08107}{{\tt
  1802.08107}}].

\bibitem{Bogers2018b}
M.~P. Bogers and T.~Brauner, \emph{{Lie-algebraic classification of effective
  theories with enhanced soft limits}},
  \href{http://dx.doi.org/10.1007/JHEP05(2018)076}{\emph{JHEP} {\bf 05} (2018)
  076}, [\href{http://arxiv.org/abs/1803.05359}{{\tt 1803.05359}}].

\bibitem{Roest2019a}
D.~Roest, D.~Stefanyszyn and P.~Werkman, \emph{{An algebraic classification of
  exceptional EFTs}},
  \href{http://dx.doi.org/10.1007/JHEP08(2019)081}{\emph{JHEP} {\bf 08} (2019)
  081}, [\href{http://arxiv.org/abs/1903.08222}{{\tt 1903.08222}}].

\bibitem{Nicolis2009a}
A.~Nicolis, R.~Rattazzi and E.~Trincherini, \emph{{Galileon as a local
  modification of gravity}},
  \href{http://dx.doi.org/10.1103/PhysRevD.79.064036}{\emph{Phys. Rev.} {\bf
  D79} (2009) 064036}, [\href{http://arxiv.org/abs/0811.2197}{{\tt
  0811.2197}}].

\bibitem{Rham2014a}
C.~de~Rham, M.~Fasiello and A.~J. Tolley, \emph{{Galileon Duality}},
  \href{http://dx.doi.org/10.1016/j.physletb.2014.03.061}{\emph{Phys. Lett.}
  {\bf B733} (2014) 46--51}, [\href{http://arxiv.org/abs/1308.2702}{{\tt
  1308.2702}}].

\bibitem{Rham2014b}
C.~de~Rham, L.~Keltner and A.~J. Tolley, \emph{{Generalized galileon duality}},
  \href{http://dx.doi.org/10.1103/PhysRevD.90.024050}{\emph{Phys. Rev.} {\bf
  D90} (2014) 024050}, [\href{http://arxiv.org/abs/1403.3690}{{\tt
  1403.3690}}].

\bibitem{Kampf2014a}
K.~Kampf and J.~{Novotn\'y}, \emph{{Unification of Galileon dualities}},
  \href{http://dx.doi.org/10.1007/JHEP10(2014)006}{\emph{JHEP} {\bf 10} (2014)
  006}, [\href{http://arxiv.org/abs/1403.6813}{{\tt 1403.6813}}].

\bibitem{Horava2009a}
P.~Ho\v{r}ava, \emph{{Quantum Gravity at a Lifshitz Point}},
  \href{http://dx.doi.org/10.1103/PhysRevD.79.084008}{\emph{Phys. Rev.} {\bf
  D79} (2009) 084008}, [\href{http://arxiv.org/abs/0901.3775}{{\tt
  0901.3775}}].

\bibitem{Horava2009b}
P.~Ho\v{r}ava, \emph{{Spectral Dimension of the Universe in Quantum Gravity at
  a Lifshitz Point}},
  \href{http://dx.doi.org/10.1103/PhysRevLett.102.161301}{\emph{Phys. Rev.
  Lett.} {\bf 102} (2009) 161301}, [\href{http://arxiv.org/abs/0902.3657}{{\tt
  0902.3657}}].

\bibitem{Andringa2012a}
R.~Andringa, E.~Bergshoeff, J.~Gomis and M.~de~Roo, \emph{{'Stringy'
  Newton-Cartan Gravity}},
  \href{http://dx.doi.org/10.1088/0264-9381/29/23/235020}{\emph{Class. Quant.
  Grav.} {\bf 29} (2012) 235020}, [\href{http://arxiv.org/abs/1206.5176}{{\tt
  1206.5176}}].

\bibitem{Harmark2017a}
T.~Harmark, J.~Hartong and N.~A. Obers, \emph{{Nonrelativistic strings and
  limits of the AdS/CFT correspondence}},
  \href{http://dx.doi.org/10.1103/PhysRevD.96.086019}{\emph{Phys. Rev.} {\bf
  D96} (2017) 086019}, [\href{http://arxiv.org/abs/1705.03535}{{\tt
  1705.03535}}].

\bibitem{Kluson2018a}
J.~Kluso{\v n}, \emph{{Remark About Non-Relativistic String in Newton-Cartan
  Background and Null Reduction}},
  \href{http://dx.doi.org/10.1007/JHEP05(2018)041}{\emph{JHEP} {\bf 05} (2018)
  041}, [\href{http://arxiv.org/abs/1803.07336}{{\tt 1803.07336}}].

\bibitem{Pajer2020a}
E.~Pajer, D.~Stefanyszyn and J.~{Supe\lt}, \emph{{The Boostless Bootstrap:
  Amplitudes without Lorentz boosts}},
  \href{http://dx.doi.org/10.1007/JHEP12(2020)198}{\emph{JHEP} {\bf 12} (2020)
  198}, [\href{http://arxiv.org/abs/2007.00027}{{\tt 2007.00027}}].

\bibitem{Bacry1968a}
H.~Bacry and J.~L{\'e}vy-Leblond, \emph{{Possible kinematics}},
  \href{http://dx.doi.org/10.1063/1.1664490}{\emph{J. Math. Phys.} {\bf 9}
  (1968) 1605--1614}.

\bibitem{FigueroaOFarrill2017a}
J.~Figueroa-O'Farrill, \emph{{Classification of kinematical Lie algebras}},
  \href{http://arxiv.org/abs/1711.05676}{{\tt 1711.05676}}.

\bibitem{FigueroaOFarrill2018a}
J.~M. Figueroa-O'Farrill, \emph{{Kinematical Lie algebras via deformation
  theory}}, \href{http://dx.doi.org/10.1063/1.5016288}{\emph{J. Math. Phys.}
  {\bf 59} (2018) 061701}, [\href{http://arxiv.org/abs/1711.06111}{{\tt
  1711.06111}}].

\bibitem{Grosvenor2018a}
K.~T. Grosvenor, J.~Hartong, C.~Keeler and N.~A. Obers, \emph{{Homogeneous
  nonrelativistic geometries as coset spaces}},
  \href{http://dx.doi.org/10.1088/1361-6382/aad0f9}{\emph{Class. Quant. Grav.}
  {\bf 35} (2018) 175007}, [\href{http://arxiv.org/abs/1712.03980}{{\tt
  1712.03980}}].

\bibitem{Endlich2014a}
S.~Endlich, A.~Nicolis and R.~Penco, \emph{{Ultraviolet completion without
  symmetry restoration}},
  \href{http://dx.doi.org/10.1103/PhysRevD.89.065006}{\emph{Phys. Rev.} {\bf
  D89} (2014) 065006}, [\href{http://arxiv.org/abs/1311.6491}{{\tt
  1311.6491}}].

\bibitem{Volkov1973a}
D.~V. Volkov, \emph{{Phenomenological Lagrangians}}, {\emph{Fiz. Elem. Chast.
  Atom. Yadra} {\bf 4} (1973) 3--41}.

\bibitem{Ogievetsky1974a}
V.~Ogievetsky, \emph{{Nonlinear realizations of internal and spacetime
  symmetries}}, {\emph{Acta Universitatis Wratislaviensis} {\bf 207} (1974)
  117--141}.

\bibitem{Witten1983a}
E.~Witten, \emph{{Global aspects of current algebra}},
  \href{http://dx.doi.org/10.1016/0550-3213(83)90063-9}{\emph{Nucl. Phys.} {\bf
  B223} (1983) 422--432}.

\bibitem{D'Hoker1994a}
E.~D'Hoker and S.~Weinberg, \emph{{General effective actions}},
  \href{http://dx.doi.org/10.1103/PhysRevD.50.R6050}{\emph{Phys. Rev.} {\bf
  D50} (1994) 6050--6053}, [\href{http://arxiv.org/abs/hep-ph/9409402}{{\tt
  hep-ph/9409402}}].

\bibitem{D'Hoker1995b}
E.~D'Hoker, \emph{{Invariant effective actions, cohomology of homogeneous
  spaces and anomalies}},
  \href{http://dx.doi.org/10.1016/0550-3213(95)00265-T}{\emph{Nucl. Phys.} {\bf
  B451} (1995) 725--748}, [\href{http://arxiv.org/abs/hep-th/9502162}{{\tt
  hep-th/9502162}}].

\bibitem{Davighi2018a}
J.~Davighi and B.~Gripaios, \emph{{Homological classification of topological
  terms in sigma models on homogeneous spaces}},
  \href{http://dx.doi.org/10.1007/JHEP11(2018)143,
  10.1007/JHEP09(2018)155}{\emph{JHEP} {\bf 09} (2018) 155},
  [\href{http://arxiv.org/abs/1803.07585}{{\tt 1803.07585}}].

\bibitem{Goon2012a}
G.~Goon, K.~Hinterbichler, A.~Joyce and M.~Trodden, \emph{{Galileons as
  Wess-Zumino Terms}},
  \href{http://dx.doi.org/10.1007/JHEP06(2012)004}{\emph{JHEP} {\bf 06} (2012)
  004}, [\href{http://arxiv.org/abs/1203.3191}{{\tt 1203.3191}}].

\bibitem{Azcarraga1998b}
J.~A. de~Azc\'arraga, A.~J. Macfarlane, A.~J. Mountain and J.~C. P\'erez~Bueno,
  \emph{{Invariant tensors for simple groups}},
  \href{http://dx.doi.org/10.1016/S0550-3213(97)00609-3}{\emph{Nucl. Phys.}
  {\bf B510} (1998) 657--687},
  [\href{http://arxiv.org/abs/physics/9706006}{{\tt physics/9706006}}].

\bibitem{Nielsen1976a}
H.~B. Nielsen and S.~Chadha, \emph{{On how to count Goldstone bosons}},
  \href{http://dx.doi.org/10.1016/0550-3213(76)90025-0}{\emph{Nucl. Phys.} {\bf
  B105} (1976) 445--453}.

\bibitem{Watanabe2011a}
H.~Watanabe and T.~Brauner, \emph{{On the number of Nambu-Goldstone bosons and
  its relation to charge densities}},
  \href{http://dx.doi.org/10.1103/PhysRevD.84.125013}{\emph{Phys. Rev.} {\bf
  D84} (2011) 125013}, [\href{http://arxiv.org/abs/1109.6327}{{\tt
  1109.6327}}].

\bibitem{Watanabe2012b}
H.~Watanabe and H.~Murayama, \emph{{Unified Description of Nambu-Goldstone
  Bosons without Lorentz Invariance}},
  \href{http://dx.doi.org/10.1103/PhysRevLett.108.251602}{\emph{Phys. Rev.
  Lett.} {\bf 108} (2012) 251602}, [\href{http://arxiv.org/abs/1203.0609}{{\tt
  1203.0609}}].

\bibitem{Hidaka2013b}
Y.~Hidaka, \emph{{Counting rule for Nambu-Goldstone modes in nonrelativistic
  systems}},
  \href{http://dx.doi.org/10.1103/PhysRevLett.110.091601}{\emph{Phys. Rev.
  Lett.} {\bf 110} (2013) 091601}, [\href{http://arxiv.org/abs/1203.1494}{{\tt
  1203.1494}}].

\bibitem{Hinterbichler2014a}
K.~Hinterbichler and A.~Joyce, \emph{{Goldstones with Extended Shift
  Symmetries}}, \href{http://dx.doi.org/10.1142/S0218271814430019}{\emph{Int.
  J. Mod. Phys.} {\bf D23} (2014) 1443001},
  [\href{http://arxiv.org/abs/1404.4047}{{\tt 1404.4047}}].

\bibitem{Griffin2015a}
T.~Griffin, K.~T. Grosvenor, P.~{Ho\v{r}ava} and Z.~Yan, \emph{{Scalar Field
  Theories with Polynomial Shift Symmetries}},
  \href{http://dx.doi.org/10.1007/s00220-015-2461-2}{\emph{Commun. Math. Phys.}
  {\bf 340} (2015) 985--1048}, [\href{http://arxiv.org/abs/1412.1046}{{\tt
  1412.1046}}].

\bibitem{Brauner2014b}
T.~Brauner and S.~Moroz, \emph{{Topological interactions of Nambu-Goldstone
  bosons in quantum many-body systems}},
  \href{http://dx.doi.org/10.1103/PhysRevD.90.121701}{\emph{Phys. Rev.} {\bf
  D90} (2014) 121701}, [\href{http://arxiv.org/abs/1405.2670}{{\tt
  1405.2670}}].

\bibitem{Rham2010a}
C.~de~Rham and A.~J. Tolley, \emph{{DBI and the Galileon reunited}},
  \href{http://dx.doi.org/10.1088/1475-7516/2010/05/015}{\emph{JCAP} {\bf 05}
  (2010) 015}, [\href{http://arxiv.org/abs/1003.5917}{{\tt 1003.5917}}].

\bibitem{Brauner2014c}
T.~Brauner, S.~Endlich, A.~Monin and R.~Penco, \emph{{General coordinate
  invariance in quantum many-body systems}},
  \href{http://dx.doi.org/10.1103/PhysRevD.90.105016}{\emph{Phys. Rev.} {\bf
  D90} (2014) 105016}, [\href{http://arxiv.org/abs/1407.7730}{{\tt
  1407.7730}}].

\bibitem{Hinterbichler2015a}
K.~Hinterbichler and A.~Joyce, \emph{{Hidden symmetry of the Galileon}},
  \href{http://dx.doi.org/10.1103/PhysRevD.92.023503}{\emph{Phys. Rev.} {\bf
  D92} (2015) 023503}, [\href{http://arxiv.org/abs/1501.07600}{{\tt
  1501.07600}}].

\bibitem{Novotny2017a}
J.~Novotn{\'y}, \emph{{Geometry of special Galileons}},
  \href{http://dx.doi.org/10.1103/PhysRevD.95.065019}{\emph{Phys. Rev.} {\bf
  D95} (2017) 065019}, [\href{http://arxiv.org/abs/1612.01738}{{\tt
  1612.01738}}].

\bibitem{Weinberg1995a}
S.~Weinberg, \emph{The Quantum Theory of Fields}, vol.~I.
\newblock Cambridge University Press, Cambridge, UK, 1995.

\bibitem{Greiter1989a}
M.~Greiter, F.~Wilczek and E.~Witten, \emph{Hydrodynamic relations in
  superconductivity},
  \href{http://dx.doi.org/10.1142/S0217984989001400}{\emph{Mod. Phys. Lett.}
  {\bf B03} (1989) 903--918}.

\bibitem{Son2006a}
D.~T. Son and M.~Wingate, \emph{{General coordinate invariance and conformal
  invariance in nonrelativistic physics: Unitary Fermi gas}},
  \href{http://dx.doi.org/10.1016/j.aop.2005.11.001}{\emph{Ann. Phys.} {\bf
  321} (2006) 197--224}, [\href{http://arxiv.org/abs/cond-mat/0509786}{{\tt
  cond-mat/0509786}}].

\end{thebibliography}\endgroup

\end{document}